%% file: bxmuonveto.tex
\pdfoutput=1
\documentclass[a4paper, 11pt]{JINST}

\usepackage{textcomp}
\usepackage{amsmath}
\usepackage{lineno}

\include{defines}

\title{Muon and Cosmogenic Neutron Detection in Borexino}

\author{
G.~Bellini$^{1}$, J.~Benziger$^{6}$, D.~Bick$^{17}$, S.~Bonetti$^{1}$, M.~Buizza Avanzini$^{1}$, B.~Caccianiga$^{1}$, 
L.~Cadonati$^{16}$, F.~Calaprice$^{4}$, C.~Carraro$^{7}$, A.~Chavarria$^{4}$, A.~Chepurnov$^{18}$,
D.~D{\textquoteright}Angelo$^{1}$, S.~Davini$^{7}$, A.~Derbin$^{9}$, A.~Etenko$^{10}$, 
F. von Feilitzsch$^{8}$, K.~Fomenko$^{11}$, D.~Franco$^{5}$, C.~Galbiati$^{4}$, S.~Gazzana$^{2}$, C.~Ghiano$^{2}$,
M.~Giammarchi$^{1}$, M.~Goeger-Neff$^{8}$, A.~Goretti$^{4}$, E.~Guardincerri$^{7}$, S.~Hardy$^{3}$, 
Aldo Ianni$^{2}$, Andrea Ianni$^{4}$, M.~Joyce$^{3}$, V.~Kobychev$^{19}$, Y.~Koshio$^{2}$, D.~Korablev$^{11}$, G.~Korga$^{13,2}$, 
D.~Kryn$^{5}$, M.~Laubenstein$^{2}$, C.~Lendvai$^{8}$, T.~Lewke$^{8}$, E.~Litvinovich$^{10}$, 
B.~Loer$^{4}$, F.~Lombardi$^{2}$, P.~Lombardi$^{1}$, L.~Ludhova$^{1}$, I.~Machulin$^{10}$, S.~Manecki$^{3}$, 
W.~Maneschg$^{12}$, G.~Manuzio$^{7}$, Q.~Meindl$^{8}$, E.~Meroni$^{1}$, L.~Miramonti$^{1}$, M.~Misiaszek$^{14}$, 
D.~Montanari$^{2,4}$, V.~Muratova$^{9}$, L.~Oberauer$^{8}$, M.~Obolensky$^{5}$, F.~Ortica$^{15}$, M.~Pallavicini$^{7}$, 
L.~Papp$^{13,2,3}$, L.~Perasso$^{1}$, S.~Perasso$^{7}$, A.~Pocar$^{16}$, R.S.~Raghavan$^{3}$, G.~Ranucci$^{1}$, 
A.~Razeto$^{2}$, A.~Re$^{1}$, A.~Romani$^{15}$, D.~Rountree$^{3}$, A.~Sabelnikov$^{10}$, R.~Saldanha$^{4}$, 
C.~Salvo$^{7}$, S.~Sch\"onert$^{12}$, H.~Simgen$^{12}$, M.~Skorokhvatov$^{10}$, O.~Smirnov$^{11}$, A.~Sotnikov$^{11}$, 
S.~Sukhotin$^{10}$, Y.~Suvorov$^{2}$, R.~Tartaglia$^{2}$, G.~Testera$^{7}$, D.~Vignaud$^{5}$, R.B.~Vogelaar$^{3}$, 
J.~Winter$^{8}$, M.~Wojcik$^{14}$, A.~Wright$^{4}$, M.~Wurm$^{8}$, J.~Xu$^{4}$, O.~Zaimidoroga$^{11}$, S.~Zavatarelli$^{7}$,
G.~Zuzel$^{14}$\\
\center{(BOREXino Collaboration)}

$^1$Dipartimento di Fisica, Universit\`a di Milano and INFN Milano, via Celoria 16, \\ I-20133 Milano, Italy

$^2$Laboratori Nazionali del Gran Sasso, SS 17bis Km 18+910, I-67010 Assergi (AQ), Italy

$^3$Physics Department, Robeson Hall, Virginia Polytechnic Institute and State University, Blacksburg, VA 24061-0435, USA 

$^4$Department of Physics, Princeton University, Jadwin Hall, Washington Road, Princeton, \\ NJ 08544-0708, USA

$^5$Astroparticule et Cosmologie APC, 10 rue Alice Domon et L\'eonie Duquet, \\ 75205 Paris cedex 13, France

$^6$Department of Chemical Engineering, Princeton University, Engineering Quadrangle, Princeton, NJ 08544-5263, USA

$^7$Dipartimento di Fisica, Universit\`a di Genova and INFN Genova, via Dodecaneso 33, \\ I-16146 Genova, Italy

$^8$Technische Universit\"at M\"unchen, James Franck Strasse E15, D-85747 Garching, Germany

$^9$St. Petersburg Nuclear Physics Institute, Gatchina, Russianon

$^{10}$RRC Kurchatov Institute, Kurchatov Sq. 1, 123182 Moscow, Russia

$^{11}$JINR, Joliot Curie str. 6, 141980 Dubna, Russia

$^{12}$Max-Planck-Institut f\"ur Kernphysik, Postfach 103 980, D-69029 Heidelberg, Germany

$^{13}$KFKI-RMKI, 1121 Budapest, Hungary

$^{14}$Institute of Physics, Jagiellonian University, ul. Reymonta 4, PL-30059 Krakow, Poland

$^{15}$Dipartimento di Chimica, Universit\`a di Perugia and INFN Perugia, via Elce di Sotto 8, I-06123 Perugia, Italy

$^{16}$Physics Department, University of Massachusetts, Amherst, MA 01003, USA

$^{17}$Institut f\"ur Experimentalphysik, Universit\"at Hamburg, 22761 Hamburg, Germany

$^{18}$Institute of Nuclear Physics, Lomonosov Moscow State University, 119899, Moscow, Russia

$^{19}$Institute for Nuclear Research, 03680 Kiev, Ukraine
}

\abstract{

Borexino, a liquid scintillator detector at LNGS, is designed for the detection of neutrinos and antineutrinos from the Sun, supernovae, nuclear reactors, and the Earth. The feeble nature of these signals requires a strong suppression of backgrounds below a few MeV. Very low intrinsic radiogenic contamination of all detector components needs to be accompanied by the efficient identification of muons and of muon-induced backgrounds. Muons produce unstable nuclei by spallation processes along their trajectory through the detector whose decays can mimic the expected signals; for isotopes with half-lives longer than a few seconds, the dead time induced by a muon-related veto becomes unacceptably long, unless its application can be restricted to a sub-volume along the muon track. Consequently, not only the identification of muons with very high efficiency but also a precise reconstruction of their tracks is of primary importance for the physics program of the experiment. The Borexino inner detector is surrounded by an outer water-\v{C}erenkov detector that plays a fundamental role in accomplishing this task. The detector design principles and their implementation are described. The strategies adopted to identify muons are reviewed and their efficiency is evaluated. The overall muon veto efficiency is found to be 99.992\,\% or better. Ad-hoc track reconstruction algorithms developed are presented. Their performance is tested against muon events of known direction such as those from the CNGS neutrino beam, test tracks available from a dedicated External Muon Tracker and cosmic muons whose angular distribution reflects the local overburden profile. The achieved angular resolution is $\sim$3$^{\circ}$-5$^{\circ}$ and the lateral resolution is $\sim$35-50\,cm, depending on the impact parameter of the crossing muon. The methods implemented to efficiently tag cosmogenic neutrons are also presented.
}


\begin{document}


\section{Introduction}
\label{sec:intro} 

Borexino is a large-volume liquid-scintillator detector for low-energy neutrinos and anti-neutrinos coming from the Sun, 
from supernovas, from nuclear reactors, and from the Earth (geo-neutrinos). 
The very low radioactive background conditions achieved for the scintillator allowed us to widen the scope of the experiment 
beyond the measurement of the  monoenergetic solar $^7$Be neutrinos \cite{bx08be7}:
measurements of  solar $^8$B neutrinos \cite{bx10b8} and of geoneutrinos \cite{bx10geo} have recently been published. 
It is planned in the future to look also for solar pp, pep, and CNO neutrinos. 

A key element to reach the required sensitivity for these low-rate and low-energy event searches is the suppression 
of the background events created by natural radioactivity and cosmic muon radiation, so muon identification and tracking are therefore crucial.

The Borexino detector was designed to have very low intrinsic background.  
The central scintillation volume, 278\,t
of ultra-pure PC (pseudocumene) doped with 1.5\,g/l of the fluor PPO (2,5-diphenyloxazole), is contained in a spherical
Inner Vessel (IV), 8.5\,m in diameter, made of 125\,\textmu m thick nylon. 
It is floating in two buffer layers consisting of PC and a small amount of the light quencher DMP (dimethylphthalate). 
These inactive shells provide shielding from long-range $\gamma$-rays and Rn emanating from external parts of the detector.
The surrounding Stainless Steel Sphere (SSS) of 13.7\,m diameter holds 2212 inward-facing 8'' photomultiplier tubes (PMT) 
that detect the scintillation light from the central region. 
All these components form the Inner Detector (ID) \cite{bx08det}.
Due to the radiopurity of all materials and the self-shielding capability of the liquid scintillator, 
background levels corresponding to rates of less than one count per ton-day 
have been achieved in the central part of the IV ($\sim$ 100\,t), the Fiducial Volume (FV). 

A second central requirement is the rejection of cosmic muons crossing the detector 
and the subsequent signals caused by muon spallation products. 
Though Borexino is located deeply underground, in Hall C of the Gran Sasso Laboratory (LNGS)\footnote{
The rock mass above the detector corresponds to about 3\,800 meters of water equivalent (mwe) 
and provides a reduction of the surface muon flux by about 6 orders of magnitude.}, 
the residual muon flux is  $\sim$\,1.2\,\textmu/m$^2$/h,
still too large for neutrino measurement, so the muons must be individually tagged. 
To accomplish this task the ID is surrounded by a powerful muon detector.
This is composed by a high domed steel tank of 18\,m diameter and 16.9\,m height filled
with 2\,100\,t of ultra-pure water and instrumented with 208 PMTs which detect the muon \v{C}erenkov emission. 
The Water Tank (WT) also serves as additional passive shielding against external radiation. 
This system is called the Outer Detector (OD). 
Moreover, the characteristic light patterns created by muons in both water 
and scintillator can be exploited to reconstruct the muon tracks. 
This is mandatory for the rejection of long-lived cosmogenic radioisotopes, such as $^{11}$C, 
that constitute the main background for the forthcoming pep and CNO neutrino detection. 
Moreover, it is an important tool for the study of cosmogenic backgrounds.

In this paper we describe the OD hardware and the identification and reconstruction of muons and of spallation neutrons.
\Sec{OutDet} gives an overview of the components of the detector and of the electronics, the data acquisition system being described in \cite{dan06phd}.
The criteria applied for the rejection of muons in both OD and ID are described in \sec{MuoVet}. 
The efficiencies of the individual methods, as well as the overall veto efficiency, 
are derived from the study of cosmic muons and of muons created by the CNGS  (CERN Neutrinos to Gran Sasso) neutrino beam \cite{web-proj-cngs}. 
In \sec{TraRec} we present three track reconstruction algorithms we have developed for crossing muons.
We evaluate their performance using muon tracks of known direction such as CNGS events observed also by the OPERA experiment, 
tracks registered by a small external auxiliary system we introduced for this purpose 
and cosmic muon and neutron samples. 
Finally, the detection of events in the wake of the passing muons is discussed in \sec{neu} which is dedicated to cosmogenic neutrons.
As the light generated by muon-induced particle showers severely interferes with the ID electronics, special techniques have been developed 
to efficiently detect both neutrons and subsequent decays of cosmogenic radioisotopes.
A brief conclusion is given in \sec{conclusion}. 
The CNGS neutrino beam and the selection of corresponding events in Borexino are described in \app{cngs}.

\section{The Outer Detector}
\label{sec:OutDet}

The OD forms the outermost layer of the Borexino detector. 
PMTs distributed on both the SSS outer surface and the OD floor register the light produced by passing muons (see \fig{bxod}). 

This section describes the hardware of the OD, in particular the reflective foils enhancing the light collection, the physical 
arrangement and encapsulation of the PMTs, their LED calibration system, the electronics read-out chain, and the digital trigger system. 

\bfig
\begin{minipage}[c]{0.59\textwidth}
\getfigw{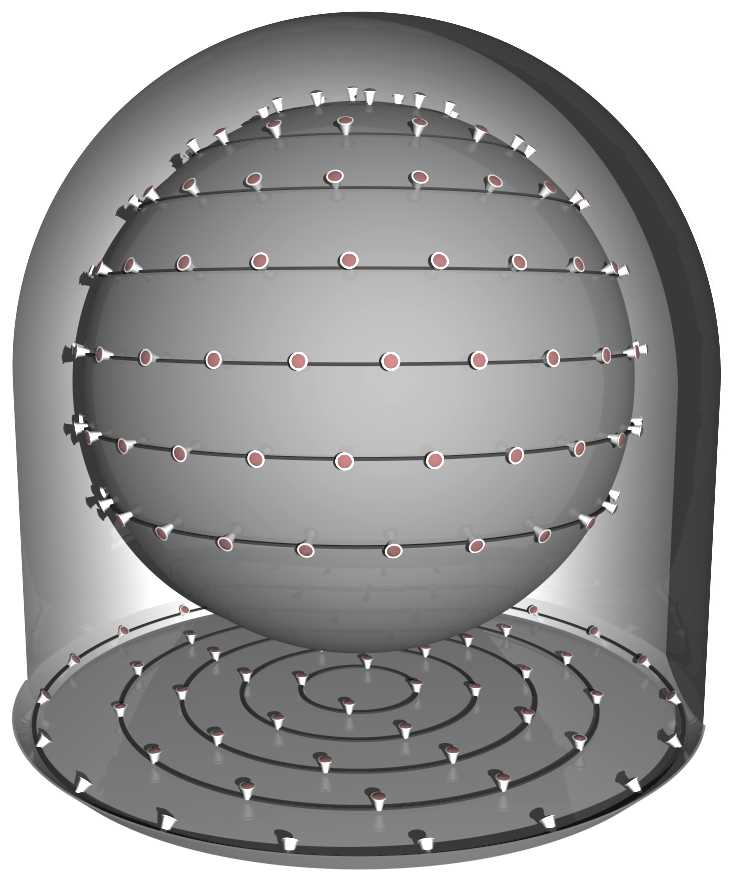}{1}
\end{minipage}
\hfill
\begin{minipage}[c]{0.39\textwidth}
\begin{tabular}{clrr}
\hline
Ring & Position & $z$\,[m] & \# of PMTs \\
\hline
+6 & SSS-UH & +7.0 & 4 \\
+5 & SSS-UH & +6.6 & 12 \\
+4 & SSS-UH & +5.6 & 17 \\
+3 & SSS-UH & +4.4 & 21 \\
+2 & SSS-UH & +2.7 & 24 \\
+1 & SSS-UH & +0.9 & 26 \\
-1 & SSS-LH & -1.0 & 26 \\
-2 & SSS-LH & - 2.8 & 24 \\
-3 & 45$^\circ$-slope & -6.8 & 20 \\
-4 & floor & -7.6 & 14 \\
-5 & floor & -7.6 & 10 \\
-6 & floor & -7.6 & 6 \\
-7 & floor & -7.6 & 4 \\
\hline
\end{tabular}
\end{minipage}
\capfig{bxod}{The Borexino Outer Detector. PMT positions are shown, grouped in rings. In the table, ring positions are indicated as stainless steel sphere (SSS) $-$ Upper (UH) and Lower (LH) Hemisphere, the OD floor and the 45$^\circ$-slope bordering the floor. The vertical $z$ coordinate corresponds to the Borexino coordinate system in which the origin is in the center of the SSS. The last column gives the number of PMTs per ring. PMT belonging to the same ring are connected by black lines in the figure (\sec{pmts}).}
\efig

\subsection{Tyvek reflector}
\label{sec:tyvek}

The steel surfaces of both the SSS and WT are covered with reflective Tyvek\footnote{Registered trademark of the DuPont company.} sheets in order to enhance the light collection (\fig{odcalib}). 
The foils are a white paper-like material, $\sim$200\,\textmu m in thickness, composed of pressed PET fibers (PolyEthylene Terephthalate). 
The spectral region relevant for  \v{C}erenkov light detection is defined both by the initial \v{C}erenkov spectrum and the quantum efficiency of the PMTs (\fig{pmt_quantum_efficiency}).
In this region the reflectivity of Tyvek is twice that of steel (\fig{tyvek_reflectivity}). Convoluting the spectrum of reflectivities with the PMT quantum efficiency and integrating returns an effective reflectivity of 38\,\% in the case of steel and 79.5\,\% for the Tyvek foils. 
Although there is a clear gain in total light collection, and therefore in veto efficiency, these foils make the muon track reconstruction in the OD more complex because of stray light (\sec{TraOD}). 

\sfig{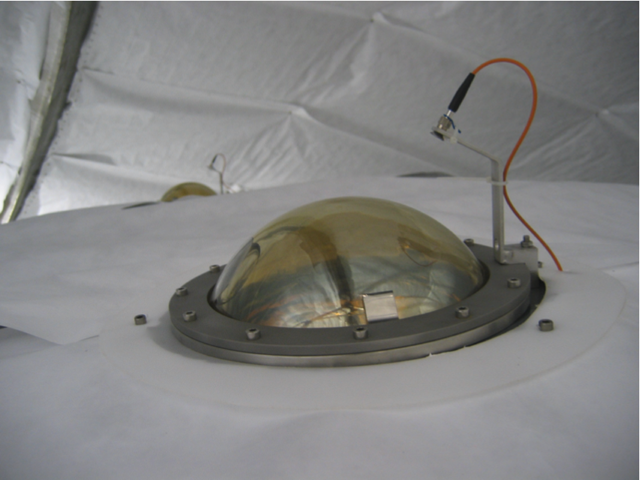}{.5}{OD PMTs. The encapsulated photocathode is emerging from the Tyvek sheets. The calibration optical fiber is also visible.}

\dfigpr{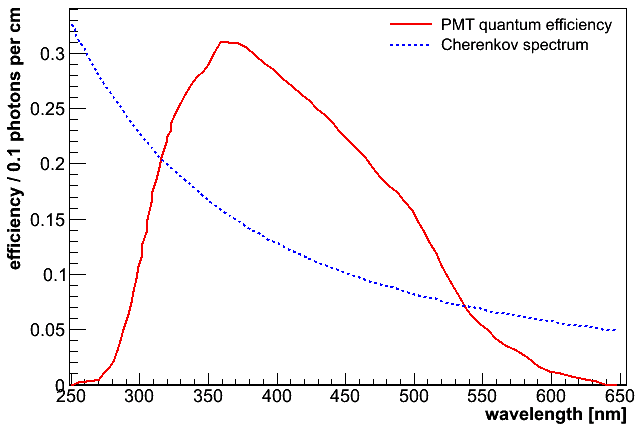}{The 8'' ETL-9351 PMT quantum efficiency and the \v{C}erenkov light emission spectrum.}
           {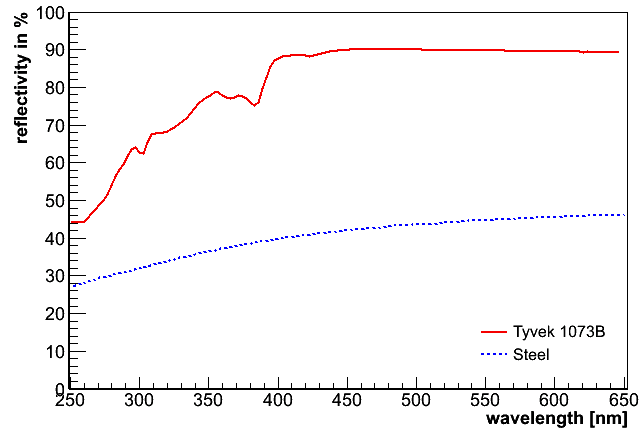}{Tyvek reflectivity as a function of wavelength compared to steel.}

\subsection{Photomultiplier tubes}
\label{sec:pmts}

The OD is equipped with 208 8'' ETL-9351 PMTs manufactured by Electron Tubes Limited (former Thorn Emi ETL) \cite{PMTs}.
Nominal operation values are listed in \tab{PmtPar}. 
The wavelength-dependent quantum efficiency is presented in \fig{pmt_quantum_efficiency}, 
where the \v{C}erenkov light spectrum is also given for comparison.

\bfig
\begin{minipage}[u]{0.49\textwidth}
\getfigw{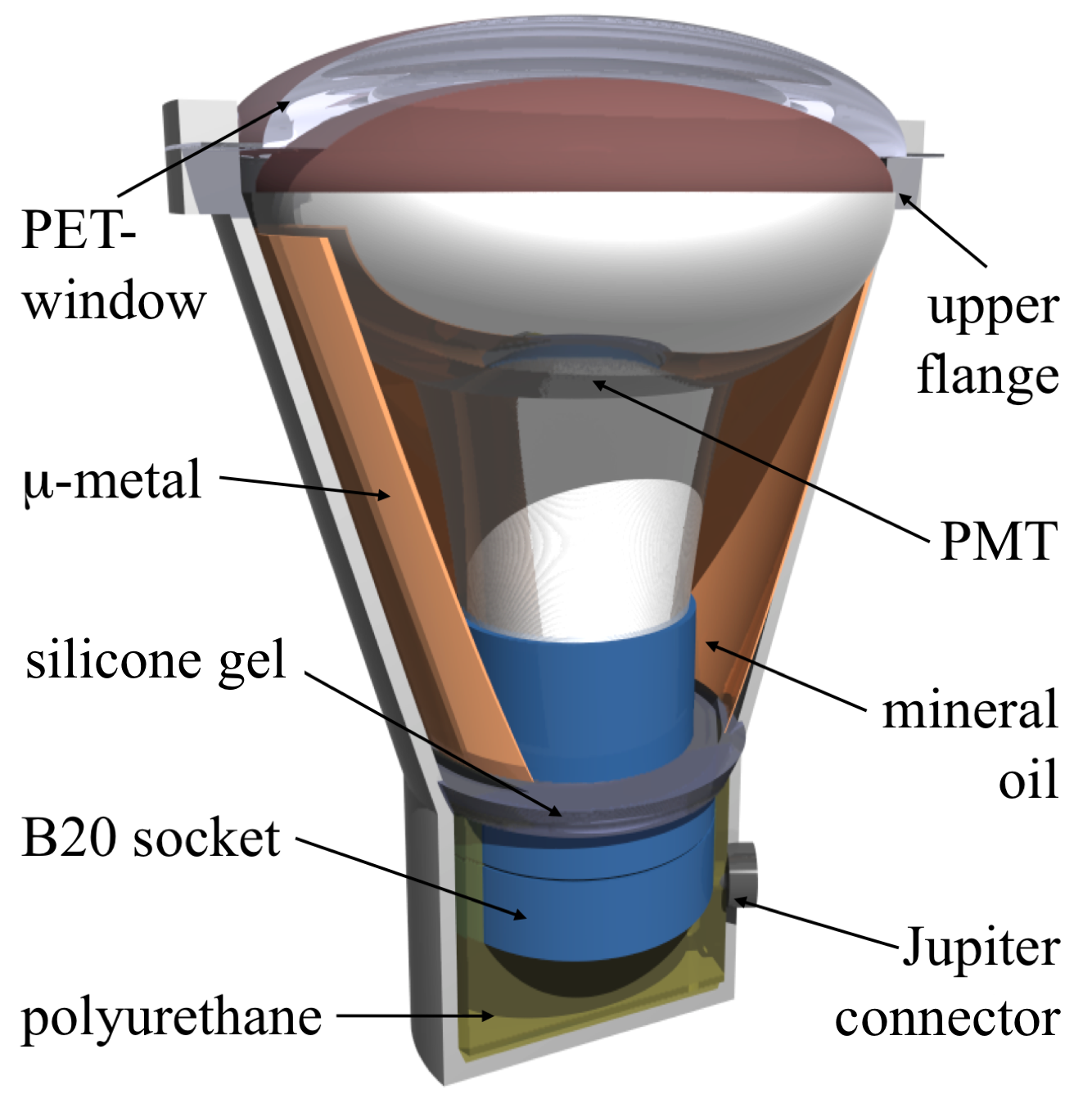}{1}
\capfig{odpmt}{Encapsulation and support structure of the OD PMTs.}
\end{minipage}
\hfill
\begin{minipage}[u]{0.49\textwidth}
\captab{PmtPar}{Manufacturer specification for 8'' ETL-9351 PMTs, operated at an amplification gain of $10^7$. \com{spe} stays for single photoelectron.}
\begin{tabular}{lr}
\hline
voltage & --1650\,V \\
\com{spe} rise time & 4\,ns \\
\com{spe} FWHM & 7\,ns \\
\com{spe} fall time & 8\,ns \\
linearity peak current & 10\,mA \\
\com{spe} peak-to-valley ratio & 2.5 \\
photocathode sensitivity (420\,nm) & 26.5\,\% \\
transit time spread (FWHM) & 2.8\,ns \\
pre-pulsing (2$\sigma$) & 3\,\% \\
after-pulsing (0.05-12.4\,\textmu s) & 2-5\,\% \\
dark counts (room temp.)& 3\,kHz \\ 
\hline
\end{tabular}
\end{minipage}
\efig

As the PMTs are operated in the aqueous environment of the OD, they are fully encapsulated to shield their base electronics from water. 
\Fig{odpmt} shows the PMT contained in a conically shaped steel case: 
a transparent ($\sim$90\,\%) PET foil at the top allows the light to reach the photocathode, retaining an acceptance angle close to 180$^\circ$. 
The foil is flexible so that the water pressure is passed on to the PMT. 
The PMT head and the dynodes structure are shielded by a conical \textmu-metal foil from Earth's magnetic field.
The remaining space between the PMT and its steel case is filled with a transparent mineral oil, 
whose refraction index of $n\approx1.5$ matches both the PET foil and the PMT glass. 
The only (unavoidable) discontinuity in the light transport arises between water ($n=1.33$) and PET foil. 
The PMT socket is plugged into the voltage divider that is in turn cast in polyurethane and in this way firmly connected to the housing bottom. 
An additional clamp (not shown in \fig{odpmt}) holds the PMT in its place\footnote{This is needed, in case the PMT is positioned upward-looking, in order to counter the buoyancy force that otherwise pulls it out of the socket.}.
Both HV and signal are fed via a Jupiter connector and a 55\,m long submarine HV cable to the outside counting room. 
PMTs mounted in the WT can be seen in \fig{odcalib}.
The design and realization of the encapsulation proved to be effective, a fact testified by the contained failure rate: 
after loosing  $\sim$2.5\,\% of PMTs during the initial filling of the WT, for the first three years of operation the failure rate has been 
only $\sim$2.5\,\%/year. 

While the original detector design foresaw an equal distribution of the 208\,PMTs on the SSS outer surface, 
rearranging the lower fourth of the PMTs to the floor of the WT significantly enhances the direct light collection for vertical muons 
and consequently the 
tracking possibilities of the detector. 
Monte Carlo (MC) simulations indicated that the veto efficiency was unaffected by the rearrangement. 
The final PMT deployment  is described in \fig{bxod}:
154 radially outward-looking PMTs on the SSS are arrayed in eight horizontal rings on the upper 2/3 of the sphere; 
34 PMTs are ordered in four concentric rings on the WT floor, 
while 20 PMTs are mounted on a  steel slope of 45$^\circ$ inward inclination that runs along the bottom of the domed steel tank; 
the slope (about 1\,m in height)  was introduced for engineering reasons. 

\subsection{Electronics}
\label{sec:elec}

Borexino PMTs produce a single photoelectron (\com{spe}) signal about 12\,mV high and 7\,ns wide (FWHM), 
corresponding to $\sim$1.6\,pC charge.
However, in the OD, a single PMT can easily see up to $\sim$100\,\com{pe} if directly hit by the \che\ cone.
An ad-hoc data acquisition chain has been designed to ensure the necessary dynamic range, without sacrificing too much the \com{spe} resolution.
The system is based on \emph{Charge-to-Time Converters} (QTCs) that integrate the PMT signals. 
The logic pulses whose duration is proportional to the input signals' area are read by commercial multi-hit \emph{Time-to-Digital Converters} (TDCs).
DAQ occurs through the VME dataway bus and is mastered by a Motorola PowerPC (PPC) computer running custom software \cite{dan06phd}.

\subsubsection{High Voltage supply and signal decoupling}
\label{sec:elec_hv}

All PMTs are operated at a nominal gain of $10^7$.
The negative operating voltage required to achieve this gain was measured for each PMT in a dark room test facility before the installation in the WT.
Typical values lie between $\sim$1150\,V  and $\sim$1750\,V and the current drawn at nominal gain averages $\sim$70\,\textmu A.
To supply this power we use a CAEN \footnote{CAEN S.p.A., Viareggio (LU), Italy; www.caen.it} SY527 mainframe with 
9 CAEN A932AP distributor boards, with 24 channels each.
The  PMTs are AC coupled, i.e. the signal and the HV travel along the same cable.
The high voltage decoupling is performed by custom passive decouplers
that also house the distribution system of calibration pulses to all channels.
After the decoupling stage, the signal features a bipolar shape, however the overshoot is very small and long ($\sim$ms) 
and its interference in the signal integration can be easily kept to a minimum.

\subsubsection{Front-end electronics}
\label{sec:elec_qtc}

\sfig{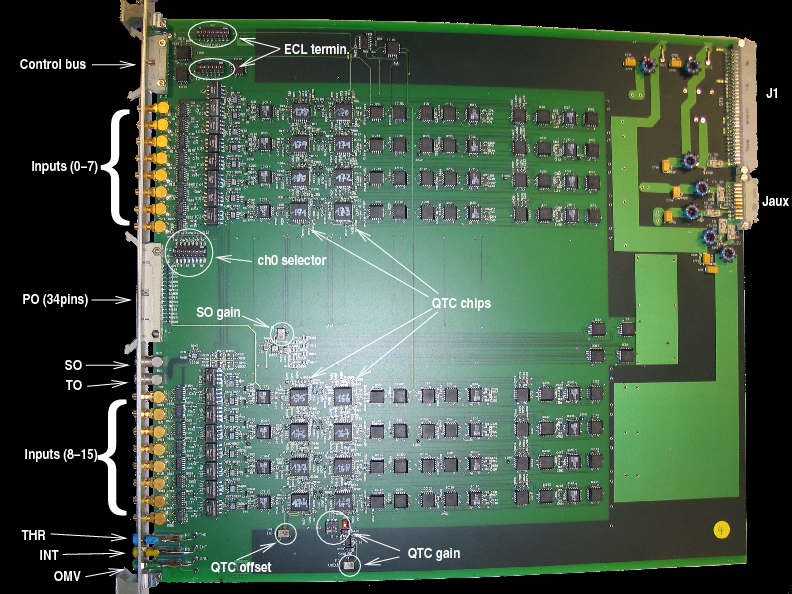}{.6}{One of the Outer Detector front-end QTC boards.}

The front-end electronics  is made of 14 QTCs, charge-to-time converter custom boards.
The board (\fig{elec_qtc_board}) is a 9U single VME unit which takes only power from the VME backplane with no connection to the dataway bus.
For each of its 16 channels the board issues a Primary Output (PO), a logic signal whose leading edge provides timing information and whose length (distance to trailing edge) is proportional to the charge.  
The board also has a Secondary Output (SO), a logic step function pulse whose height is proportional to the number of channels firing in coincidence.
SO signals are sent to the \emph{Muon Trigger Board} (MTB) for the Outer Muon Trigger (OMT) formation (\sec{elec_trg}). 
Both the width and the height of the steps can be trimmer-adjusted on the boards.
Finally, the board has a Tertiary Output (TO) issuing an analog sum of the input signals with a built-in fixed amplification factor of about two\footnote{
These outputs were introduced as an additional analog trigger system, however the performance of the digital trigger proved to be beyond expectations making the exploitation of the TO unnecessary.}.

The input signal is split in three paths reaching the QTC inputs (amplif.: $\times 5$), the threshold comparator (amplif.: $\times 11$) and the TO summing block.

The heart of the channel block is the LeCroy\footnote{LeCroy Corporation, Chestnut Ridge, NY (USA); www.lecroy.com} \com{mqt}300 QTC chip. 
The comparator is operated with a threshold of -20\,mV, corresponding to $\sim$0.2\,\com{pe};
its logic pulse gates the QTC for a time $\Delta t$, set to $\sim$110\,ns, a value optimized to preserve linearity over the whole dynamic range.
The chip integrates the input pulse together with $V_0$, its offset (or \emph{pedestal}), issuing a logic pulse whose length is given by:
\beq
\textrm{PO} \propto V_0 \cdot \Delta t + \int_{\Delta t} S(t)dt \imply \textrm{PO} = P + G \cdot Q
\eeq
where $P$ is the pedestal length  in ns and $G$ is the channel gain, trimmer-adjustable in the range [10-60]\,ns/pC.
We use the lowest possible gain that ensures a dynamic range beyond 100\,\com{pe} (corresponding to $\sim$\,2\textmu s pulse).
Similarly a trimmer allows tuning of the pedestal length: 
typical values are $\sim$700\,ns.
The comparator is disabled during the PO issuing and re-enabled immediately after. 
The dead time of the channel is therefore the time occurring between the closure of the integration gate and the end of the PO signal.

The QTC boards are daisy-chained by the control bus running along a single ribbon cable. 
The MTB is the bus master and its behavior is fully programmable through the PPC. 
The control bus includes an \com{enable} line, a test pulse, controls setting the run mode 
and integrity signals (described in \sec{elec_tdc}).

Two special run modes, enabled via control lines, allow the gate width and  pedestal length to be sent through the PO.
This is done at the beginning of every run. 
The pedestals are then subtracted during data processing and the gate widths are available for precise charge calibration.

\subsubsection{Digital electronics}
\label{sec:elec_tdc}

The digital electronics of the OD is made of two CAEN v673 TDC units.
The unit houses 4 TDC chips, each one with 32 channels, yielding an overall 128 channels,
with a time resolution of LSB $\simeq$\,1.04\,ns (\emph{Least Significant Bit}).

The TDCs are multi-hit devices with the ability to record up to 16 signal edges (rising or falling) 
for each event with a time depth programmable up to 64\,\textmu s. 
We select 16\,\textmu s to closely match the trigger gate to the ID electronics\,\cite{bx08det}.
The minimum time distance between 2 successive edges is 8\,ns, 
a requirement well met by our QTC signals, typically in the range $\sim$[0.7,2]\,\textmu s. 

The unit keeps digitizing incoming pulses but data are periodically overwritten unless a trigger occurs.
The unit works in \com{common stop} mode, with the trigger signal (\sec{elec_trg}), properly delayed, being the \com{stop}.
Each chip has 4 event memory buffers that allow deadtime-less operation but which are easily filled by a fast sequence of triggers.
\com{stop} signals arriving when the 4 buffers are full are ignored by the board.
For this reasons the ability of the DAQ to empty the buffers in the shortest possible time is particularly critical.

The TDC control bus is daisy-chained among the boards, mastered by the MTB (\sec{elec_trg}) 
and includes \com{stop} and \com{reset} signals to TDCs as well as \com{full} and \com{busy} flags from the TDCs.
When the \com{busy} condition is on either board, the MTB issues a \com{veto} to the main trigger to avoid DAQ jamming.

Synchronism among the 8 chips is not guaranteed by the units themselves: 
a \com{full} or a \com{busy} condition may regard only one or a few chips, while the others keep recording data.
Moreover a read-out operation may result in data chunks belonging to different events according to the presence or absence of data in a chip for a given event.
In our design the synchronism is insured externally by the presence of integrity channels. 
These are the first channel of each QTC board, pulsed in time with the trigger signal to guarantee a minimum of one hit per event in every chip.
In this way event buffer occupancy of the 8 chips can be described by a single variable.

Read and write operations are independent.
The first word is the value of a counter holding the number of events occurred since the last read operation.
Following hit words contain the difference of the gray counter readings upon each hit and common stop arrival, 
i.e. the negative hit time relative to the trigger signal in clock ticks.
The board supports sequential as well as block read-out as described in \cite{dan06phd}.

\subsubsection{Trigger system}
\label{sec:elec_trg}

In Borexino the trigger condition is a general condition for the whole system (\emph{Borexino General Trigger}, BGT)
which means that the whole detector is acquired simultaneously, no matter whether the trigger condition was risen by inner or outer detector.
 
\sfig{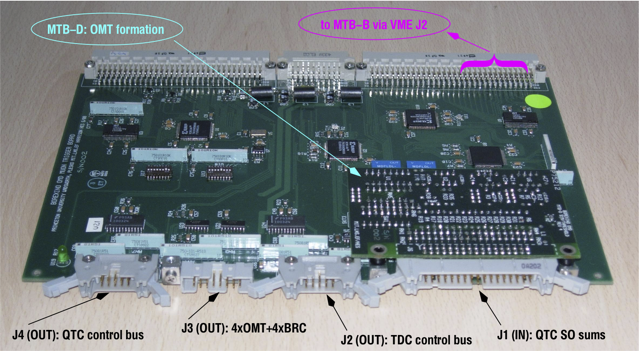}{.5}{The Muon Trigger Board (MTB), before front panel application.} 

The OD features a trigger subsystem, fully integrated in the main one which receives 
and handles the BGT and that generates the OMT signal 
and forwards it to the main trigger system.
These tasks are carried out by the MTB, a custom 6U VME unit (\fig{elec_mtb}).
The board handles both QTC and TDC control busses (\sec{elec_qtc} and \sec{elec_tdc}). 
The lines can be set or checked through VME by the PPC software.   

The board interface receives from the main trigger system 
the BGT and the \com{trgid}, a 16-bit  pattern to flag data with a universal event number.
The board in turn sends out the OMT and the OR of the two TDC \com{busy} flags.

Upon BGT arrival, the MTB issues the integrity signal to the QTCs, the \com{common stop} to the TDCs 
(delayed $\sim$70\,ns with respect to the integrity signal), acquires the \com{trgid} pattern and finally raises a VME interrupt, 
within 200\,ns of the \com{common stop} pulse, informing the DAQ that an event is ready to be read.

The task of OMT formation is delegated to a daughter board for noise isolation.
As described in \sec{elec_qtc}, the QTC boards issue (SO) a step function encoding the number of channels fired in the board. 
The width of the steps defining the trigger time window is usually set to 150\,ns, while the height is set to 100\,mV/channel.
Signals are summed and discriminated against a programmable threshold (8-bit \com{dac}), which can be set at run start through the run controller.
The inputs have a dynamic range that spans up to +2\,V before saturation occurs, allowing the threshold to be set up to almost 20 channels.
During normal data taking we use the equivalent of 6-7 PMTs (see also \sec{mtf}).

\subsection{Calibration System}
\label{sec:led}

The calibration system of the PMTs must provide both a relative time alignment with sub-ns precision 
and a charge calibration on \com{spe} level. 
The accuracy in time calibration has a large impact on the performance of the OD muon tracking (\sec{TraOD}), 
but has also influence on the efficiency of the muon identification flags (\sec{MuoTag}). 
Providing a synchronous time-signal on all photocathodes, relative offsets and jitter of both the PMT and the subsequent electronics chain can be determined. 
The calibration of the PMT \com{spe} response ensures the comparability of photon numbers detected in individual PMTs 
and is necessary for the determination of the PMT center-of-charge that is the basis of the OD tracking.

The OD calibration system is based on external LEDs. 
The light of individual LEDs is fed by 55\,m long optical fibers\footnote{
Huber-Suhner (Germany), commercial fiber, length: 55\,m, diameter: 200\,\textmu m, SMA connectors.} to the front of the OD PMTs (\fig{odcalib}). 
The UV-LEDs\footnote{Roithner Lasertechnik (Vienna, Austria), NSHU550 LED} feature a peak emission at 370\,nm  and a time jitter of 0.8\,ns. 
They are mounted on 18 customized VME boards of 12 LEDs each. 
The boards include programmable delay lines on every channel to correct inter-channel time offsets 
and allow for the possibility to adjust individually the voltage across LEDs. 
The individual LEDs are cross-calibrated using an outside PMT of the same type as the OD PMTs. 
This makes it possible to identify time and voltage parameters that achieve a synchronized light signal of equal intensity. 
This operation occurred immediately before the beginning of the data taking and could be repeated upon necessity during a maintenance shutdown.

A custom control software accessed via the run controller allows the timing and intensities to be varied during DAQ if necessary.
A weekly calibration of the OD PMTs is performed, monitoring drifts in the timing and charge of  \com{spe} signals. 

\Fig{LedCal}a shows the sum signal of $10^5$ LED calibration events, corresponding to a typical calibration run.
The solid red (black dashed) line shows the overall time distribution after (before) applying the time correction computed 
from the individual channels offset from the overall mean. 
Fitting the main peak with a Gaussian, the width improves from 3.5 to 2.1\,ns (1$\sigma$). 
This directly corresponds to the improvement in relative PMT timing.

\Fig{LedCal}b shows the \com{spe} charge spectra before (black dashed) and after (solid red) calibration. The first was corrected by the mean charge of all channels to allow a comparison of the distributions. Here, the calibration gives correction factors of 0.5 to 2 to equalize the average \com{spe} charge to a common value. As a consequence, the \com{spe} peak is more prominent after calibration.

\dfigsr{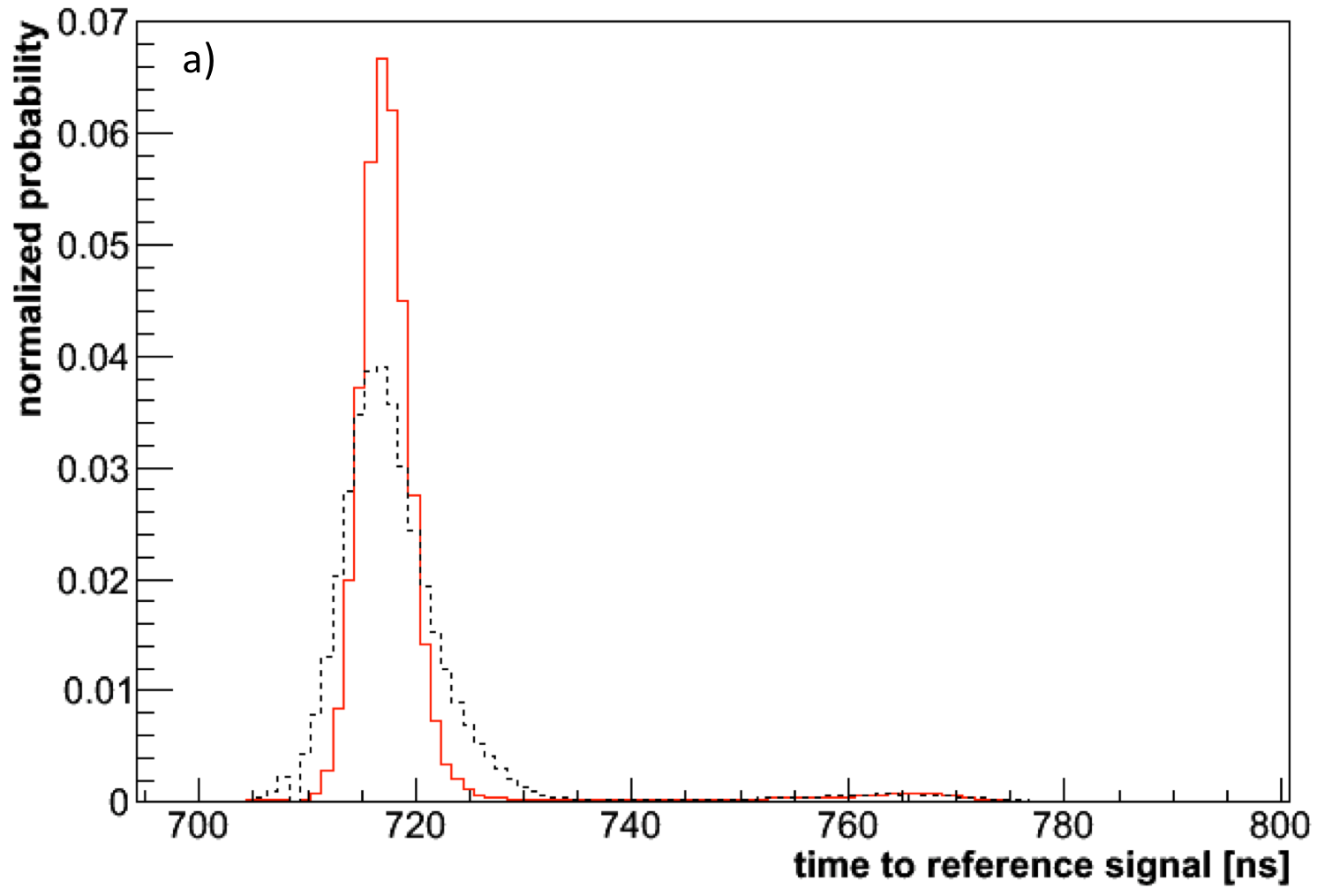}{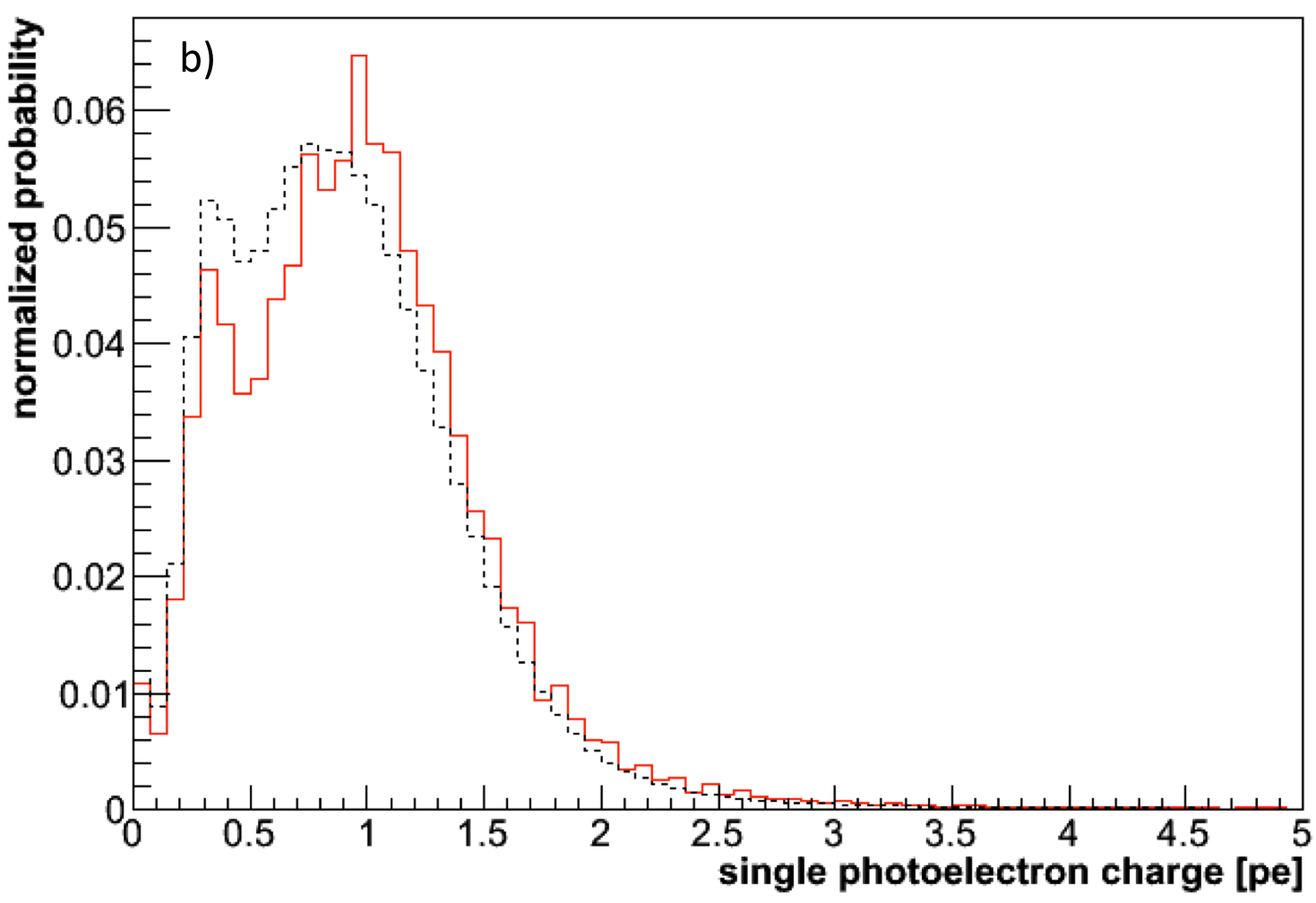}{LedCal}{Impact of LED calibration: Time (panel a)) and \com{spe} charge (panel b)) distributions  
before (black dashed line) and after (solid red line) applying calibration data.}

Evaluating the calibration runs over three years of data taking, the time offsets of individual PMTs are found to change by an average of 0.5\,ns from run to run, while the \com{spe} charge calibration shows a 10\,\% spread. The largest values observed are 4\,ns in  timing and 40\,\% in \com{spe} charge. For about 15\,\% of the channels, no calibration data is available, as the optical connection via the glass fibre is defective. These channels are not corrected for time; their \com{spe} charge is corrected by the mean value determined for all PMTs.

The use of LEDs provides a larger dynamic range of light output intensities compared to the ID system that is based on a single laser pulse distributed to the individual PMTs \cite{bx08det}. 
The OD system was designed for a charge calibration of PMTs not only in \com{spe} mode, 
but also in the region above 40 photoelectrons (\com{pe}), given the much wider dynamic range of PMT pulses expected. However, only time and \com{spe} calibration are applied in regular intervals.


\section{The Muon Veto}
\label{sec:MuoVet}

Muons are the only relevant component of cosmic radiation penetrating the rock shielding of about 3\,800\,m.w.e. over the LNGS underground Halls. 
At this depth, the muon surface flux is reduced to $\sim$1.2\,m$^{-2}$h$^{-1}$. 
Due to their mean energy of $\langle E_\mu \rangle \approx$ 270\,GeV \cite{macro02}, 
most muons travel through the detector without any significant loss of energy. 

\sfig{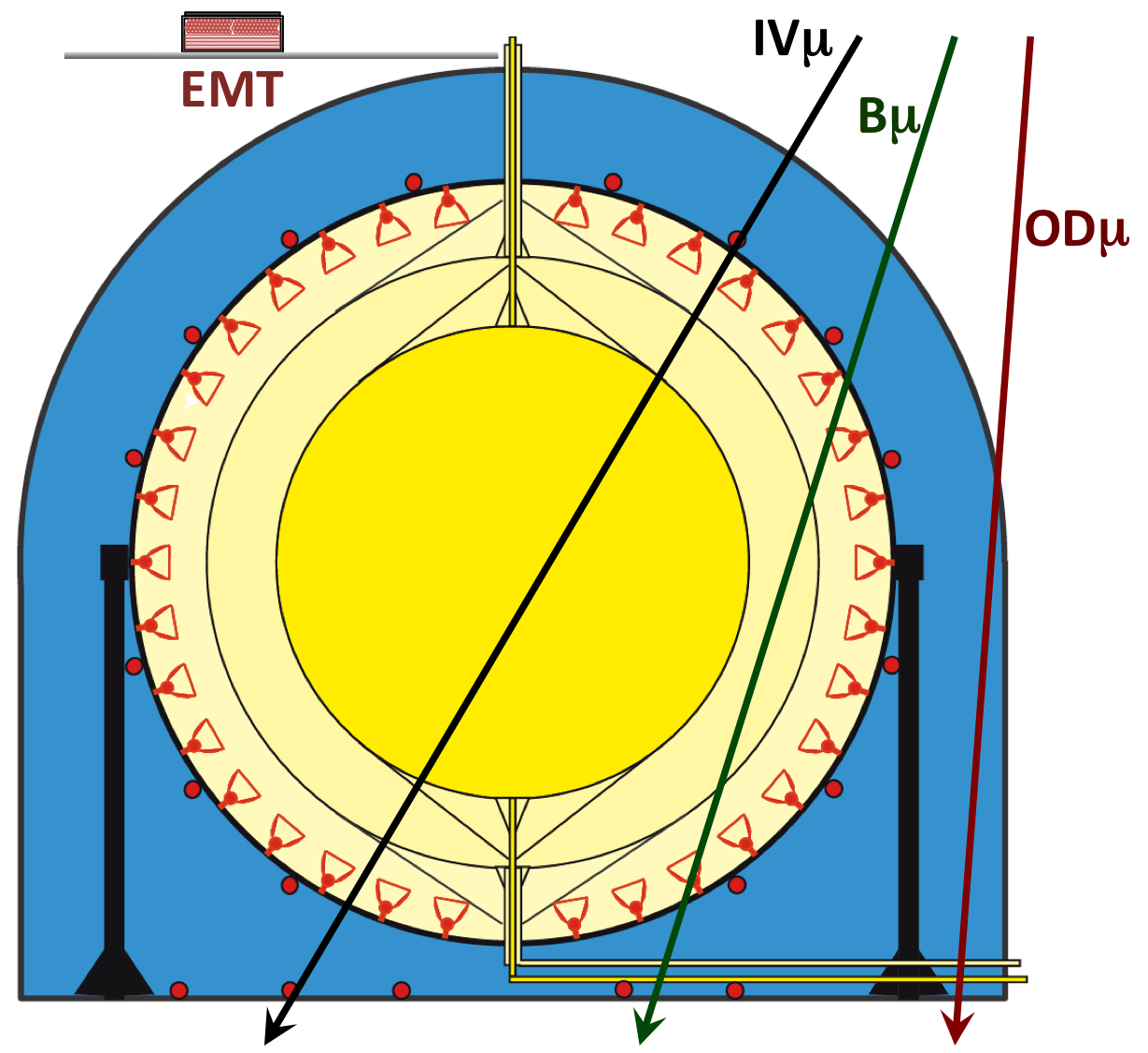}{.6}{Definition of muon subsamples. Muons crossing only the Outer Detector are abbreviated as OD\textmu's. Muons traversing the Inner Detector (ID\textmu's) are further distinguished in Inner Vessel-crossing muons (IV\textmu's) and only Buffer-crossing muons (B\textmu's). In addition, the position of the auxiliary External Muon Tracker (EMT) is shown (\ref{sec:perform_emt}).}

\Fig{muon_definitions} shows the possible scenarios of muons crossing Borexino. 
Muons passing through the scintillator of the Inner Vessel (IV\textmu's) create a very large light output as they deposit hundreds of MeV of ionization energy. 
In this case, the limitations of both the PMTs and the electronics cause a distortion of the energy information. Still, the visible energy is comparable to point-like events of several tens of MeVs. As this energy is far above the {$^7$Be}-$\nu$ shoulder or even the endpoint of the {$^8$B}-$\nu$ recoil energy spectrum, IV\textmu's are not a serious background for solar neutrino observation.
The situation changes for muons crossing the ID buffer without touching the IV. 
Buffer muons (B\textmu's) generate light partly by the \v{C}erenkov effect 
and partly by the residual scintillation of the buffer liquid, despite the DMP quencher dissolved in it \cite{bib:DMP}. 
These situations are dangerous, as the B\textmu \; light output is low enough to mimic a solar neutrino signal. 
The amount of produced light depends almost linearly on the length of the muon trajectory in the buffer. 
The corresponding muon signals cover therefore the whole range of solar neutrino recoil energies. 
The visible energy spectra of muons and point-like scintillation events are presented in \fig{nhits_pointlikes_and_muons}.
Here and in the rest of this paper, the energy variable adopted is the number of PMT hits.
The relation between this variable and the energy actually released in the detector is therefore linear only in the sub-MeV range, 
while at higher energy the effect of multiple photoelectrons being generated in the same channel and contributing to the same hit determines deviations from linearity. 
More details can be found in \cite{bx08det}.

Finally muons crossing the OD only (OD\textmu's) are acquired with a special trigger identification flag indicating that no light is present in the ID. 
Unless otherwise noted these events are not included in any analysis.

\sfig{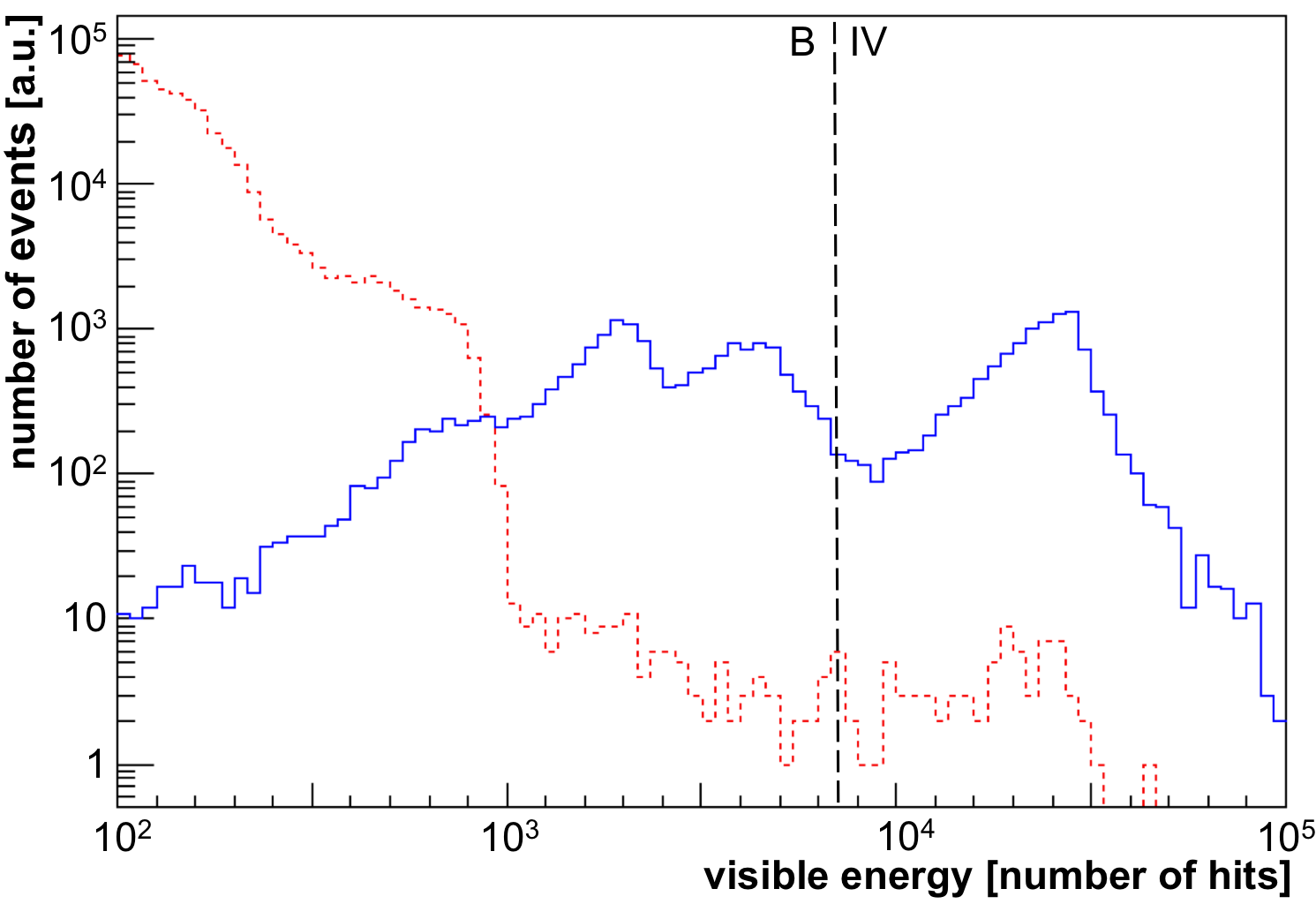}{.6}{Visible energy spectra (in number of PMT hits) for point-like (\textit{dashed red}) and muon-like (\textit{solid blue}) events in the Inner Detector, the latter flagged by the OD muon trigger (MTF).} 

The probability of a muon to stop in the detector is:
\beq
P_{\text{\textmu}}^{\text{stop}} = \frac{l^{\mathrm{av}}_{\text{\textmu}}}{\lambda}\cdot\rho_{\mathrm{PC}}  
\eeq
where $\rho_{\mathrm{PC}}=0.88 \, \mathrm{g}/\mathrm{cm}^3$ is the density of Pseudocumene,  
$\lambda\simeq700 \, \mathrm{m.w.e.}$ is the absorption length of cosmic muons for the LNGS rock coverage \cite{kud03} 
and $l^{\mathrm{av}}_{\text{\textmu}}$ is the average path length of the muons in the detector. 
Due to saturation effects following the passage of an IV\textmu, we are only able to observe muon decay of B\textmu's. 
In this case $l^{\mathrm{av}}_{\text{B\textmu}}=7.2$\,m and we obtain $P_{\text{B\textmu}}^{\text{stop}} \sim 1.0$\%. 
To check this expectation, we select B\textmu's events that feature two clusters of hits and reject those that can be ascribed to PMT after-pulses 
and other re-clustering effects in the muon signal tail, 
via the application of specifically tuned pulse shape cuts.
In the remaining sample the second cluster is therefore due to muon decay.
This is testified by the profile of the time difference between the two clusters, which correctly shows an exponential behavior. 
The fit returns $\chi^{2}/\mathrm{ndf} = 158/124$ and $\tau=(2.14\pm0.05)$\,\textmu s. 
Herein, ndf stays for the number of degrees of freedom.
An approximate evaluation of the cut efficiencies allows us to infer a rate of decaying muons of $\sim$28\,cpd, 
corresponding to $\sim$1\% of B\textmu's, in good agreement with the prediction.

\subsection{Muon Identification Criteria}
\label{sec:MuoTag}

Muon identification tags concern both the OD and the ID.
The OD muon tags search for an increase in OD PMT activity above usual dark noise fluctuations; 
they operate at two different stages of data processing, one hardware, the other software. 
For the ID we observe that typical muon pulses feature longer rise and decay times than neutrino-like scintillation events of comparable energy, 
as - depending on the track length - a muon crossing the buffer needs up to 45\,ns to deposit all its ionization energy. 
In addition, the number of detected photons is several orders of magnitude above the usual light output of point-like events, 
prolonging the signal substantially by generating large amounts of PMT after pulses. 
Pulse shape discrimination (PSD) can therefore be used to identify muons.

\subsubsection{Muon Trigger Flag (MTF)}
\label{sec:mtf}

The first tag is a hardware trigger of the OD (see \sec{elec_trg}), called the Muon Trigger Flag (MTF). 
The trigger condition is given by a significant increase in number of hits in a time span comparable to the transit time of a muon. 
The threshold is set to 6 PMTs firing within a timegate of 150\,ns. 

\subsubsection{Muon Clustering Flag (MCF)}
\label{sec:mcf}

The Muon Clustering (MCF) algorithm is an offline software trigger based on the acquired OD data. 
Its trigger condition is a slight refinement of the MTF: 
two subsets of PMTs are regarded separately, distinguishing between hits at PMTs mounted on the SSS and PMTs mounted on the floor. 
The trigger condition is met if 4 PMTs of either subset fired within 150\,ns.

\subsubsection{Inner Detector Flag (IDF)}
\label{sec:idf}
The Borexino analysis software uses a number of variables that are sensitive to either rise or decay time of the acquired pulses. 
The leading edge of the signal is described by the peak time variable $t_\mathrm{p}$, the time difference between start and peak of the pulse. 
The mean time $t_\mathrm{m}$ is defined as the average of the time differences between the start of the pulse and the individual hits in the pulse and is sensitive to the overall signal duration. 

\dfigsr{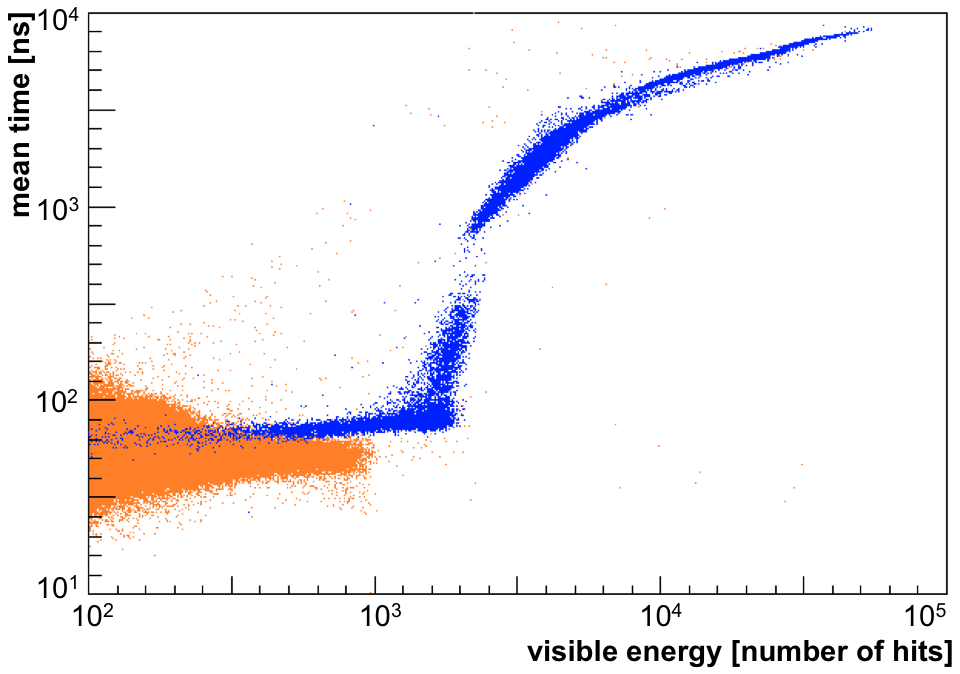}{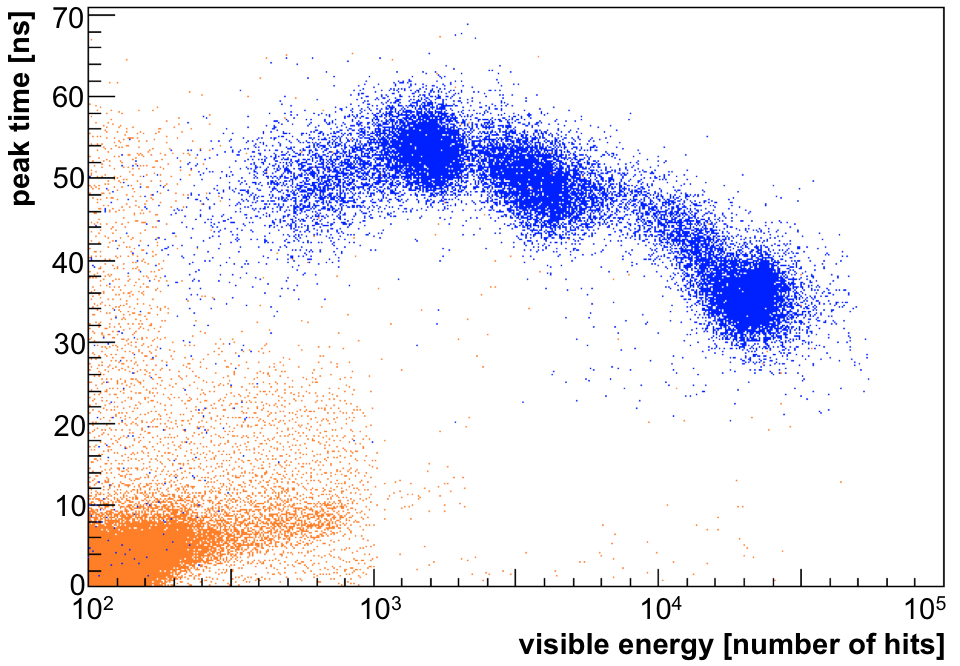}{MeaPea}{Mean time $t_\mathrm{m}$ (left panel) and peak time $t_\mathrm{p}$ (right panel) variables as a function of the visible energy of ID events. Blue dots represent events flagged by the OD tag MTF as muon, orange dots correspond to point-like events.} 

\Fig{MeaPea} shows $t_\mathrm{m}$ and $t_\mathrm{p}$ as a function of the visible event energy $E_\mathrm{vis}$. 
In the low (100$-$900 hits) and medium $E_\mathrm{vis}$ range (900$-$2\,100 hits), $t_\mathrm{p}$ is the more efficient discrimination parameter. 
The muon discrimination thresholds are chosen to be $t_\mathrm{p}$\,>\,40\,ns for the low range and $t_\mathrm{p}$\,>\,30\,ns for the medium one. 
For high energies, PMT after-pulses begin to contribute significantly to the signal, increasing especially $t_\mathrm{m}$. 
In the highest range at $E_\mathrm{vis}>2\,100$\,hits, a cut $t_\mathrm{m}>100$\,ns is applied. 
In general, the limits were chosen conservatively to avoid the false identification of point-like events as muons. 
The pulse shape Gatti parameter $g$ \footnote{The Gatti parameter \cite{gatti70} is computed from a pulse shape analysis. 
Its value is deduced from an algorithm implementing an optimum filter technique and developed for the discrimination of $\alpha$-like ($g>0$) 
and $\beta$-like ($g<0$) events. Reference shapes originating from MC studies in combination with intrinsic $\alpha$ and $\beta$ contaminants 
of the scintillator such as $^{214}$Bi and $^{214}$Po whose fast coincidence allows a very clean sample selection\cite{bx08PS}.} 
is used in the low energy region to reject scintillation events located in the most upper 
region of the Inner Buffer ($z>4$\,m) and arising from a small leak of scintillator into this volume. 
These very peripherical events show $t_\mathrm{p}$ values larger than for the other scintillation events and would therefore be identified as muons, however they are still recognized as $\beta$-like by the Gatti algorithm.
In the high energy region the Gatti parameter is instead used to reject events due to electronic noise or re-trigger on after-pulses after the actual muon. 
These events were present only in the very first months of data taking, before a hardware protection was introduced, 
and show very high $g$ values that allow an easy identification.

The criteria for muon identification are summarized in \tab{IdfDef}.

\btab
\captab{IdfDef}{Criteria for Muon Identification in the Inner Detector.}
\begin{tabular}{|c|c|c|c|}
\hline
Number of Hits & Mean Time      & Peak Time      & Pulse Shape \\
               & $t_\mathrm{m}$ & $t_\mathrm{p}$ & Gatti Parameter g\\
\hline
$100-900$      & $\times$       & $>$40\,ns      & $>$0.2 (if $z$$>$4\,m)\\
$900-2100$     & $\times$       & $>$30\,ns      & $\times$ \\
$>2100$        & $>$100\,ns     & $\times$       & $<$0.55\\
\hline
\end{tabular}
\etab

\subsection{Muon Tagging Efficiency}

The efficiencies of the available muon tags were investigated using two  approaches. 
First, samples of muon events were selected independently of any tagging flag and have been used to find directly the identification efficiencies (\sec{AbsEff}). 
Second, each flag has been tested against a muon sample selected by either of the remaining two flags (\sec{RelEff}). 

The analysis of this section focuses on ID\textmu's and includes an energy threshold of 80 hits. 
This value is not higher than the energy threshold applied in any published or foreseen Borexino analysis. 
We exploited all available data of years 2008 and 2009 except runs not passing the validation criteria \cite{wur09phd}. 
The remaining data set corresponds to an exposure of $\sim$\,500\,d including $\sim$2$\cdot$10$^{7}$ scintillation events, 
$\sim$2$\cdot$10$^{6}$ cosmic muons and $\sim$2.7$\cdot$10$^{4}$ events arising from the CNGS $\nu_{\mu}$ beam (see \app{cngs}).

In all these studies, the selected test samples contained very small (if any) impurities due to non-muon events.
When sizable, such impurities have been subtracted from the sample; however, we conservatively regard the resulting efficiencies as lower limits.

All errors reported on efficiencies throughout the section have been computed at 95\% C.L. using Clopper Pearson Intervals.

\subsubsection{Direct Efficiency Determination}
\label{sec:AbsEff}

The key requirement in selecting a test muon sample is that the flag being tested is not dependent on parameters used for the selection.
As detailed below, high-energetic ID events present in normal neutrino runs provide an appropriate test sample for the OD related flags, 
while events on time with CNGS $\nu_{\text{\textmu}}$ beam spills provide a reliable test sample for IDF.

\subsubsection*{Direct OD flags efficiency via high energy ID events}
\label{sec:VetHEM}

\Fig{nhits_pointlikes_and_muons} shows the ID energy spectra of events both flagged and not flagged by the OD MTF tag. 
As can be seen, at high visible energies, scintillation events in the ID are predominantly caused by IV\textmu's.
We selected events with more than 7\,000 hits, corresponding to a visible energy of $\sim$200\,MeV or $\sim$10$^5$ photoelectrons. 
This estimate can be derived from the fraction of after-pulses in the signal ($\sim$5\,000) and from the average probability of after-pulsing (5\,\%). 
This energy is higher than the most energetic solar neutrinos (the hep-$\nu$'s with an end-point energy of 18.8\, MeV), 
and far above the maximum energies of the radioactive decays in the natural U/Th chains ($^{208}$Tl features a Q-value of 5.0\,MeV). 

The reason why we regard this choice as valid is that the efficiencies of the OD flags are independent of the visible energy detected in the ID: 
the PMT coverage of the OD shows no relevant gaps to the cosmic muon flux; all such muons $-$ independently of the exact track geometry $-$ have
to cross several meters of water before entering the steel sphere, thereby creating comparable amounts of \v{C}erenkov light. 
The result of this study can therefore be assumed to hold over the whole energy range of ID events.

Impurities in the selected sample could be caused by fast neutrons or by ID-contained atmospheric neutrino events.
Although we cannot measure this contamination of the sample we expect such events to account for a very small fraction and we neglect them.

The resulting efficiencies are $\varepsilon_\mathrm{MTF}\ge(99.25\pm0.02)\,\%$ for MTF and $\varepsilon_\mathrm{MCF}\ge(99.28\pm0.02)\,\%$ for MCF 
(see first column of the ``Direct Efficiencies'' section in \tab{TagEff}). 

\subsubsection*{Direct ID flag efficiency via CNGS muon events}
\label{sec:VetCNGS}

The CNGS neutrino beam (see details in \app{cngs}) provides a sample of muon events
created by $\nu_{\text{\textmu}}$ charged current reactions in the upstream rock or the detector. 
As these events can be easily assigned to the beam spills via their GPS time stamp, 
they are in principle an ideal test sample for the muon tagging flags.
However, the OD is optimized for vetoing and tracking the cosmic muons which arrive preferentially around the vertical direction. 
On the contrary, the CNGS neutrino beam arrives at the detector at an angle of about $3.2^{\circ}$ below the horizon, 
and therefore generates mostly horizontal muon tracks. 
As a result, the distribution of the PMTs is not ideal (especially for muons crossing the non-instrumented lower quarter of the sphere) 
which leads to a decreased veto efficiency. 
The ID instead, due to its spherical symmetry, features an ability for muon flagging which is independent of the track direction. 
Upon these considerations, CNGS muon events are not a good test sample for OD-related flags, while they are perfectly valid for IDF.

A contamination of the sample arises from {$^{14}$C} and {$^{210}$Po} 
decay events which can occur in accidental coincidence with the beam spills. 
In \app{cngs} we have computed that $<0.06\%$ of the sample events considered here can be regarded as accidental coincidences. 
Although it is a very small effect, these events have been subtracted from the test sample.

We have tested the IDF tag in four energy regions.
The results are shown in the second column of the ``Direct Efficiencies'' section of \tab{TagEff}. 
The general increase of efficiency with $E_\mathrm{vis}$ is expected 
as the larger number of registered PMT hits improves the temporal pulse shape resolution. 
The global energy-independent efficiency was determined to be $\ge95.86^{+0.24}_{-0.25}\%$.

\btab
\captab{TagEff}{Efficiencies of the available muon identification tags. 
All numbers are lower limits as the presence of non muon events in the test samples cannot be excluded. 
Errors not shown if smaller than $10^{-4}$.}
\begin{tabular}{|lc|cc|ccc|}
\hline
Tag			   & $E_\mathrm{vis}$ & \multicolumn{2}{c|}{Direct Efficiencies}& \multicolumn{3}{c|}{Mutual Efficiencies}                       \\
                           & [hits] 	      & vs. High E & vs. CNGS                   & vs. IDF   & vs. MTF     & vs. MCF                              \\
\hline
$\varepsilon_\mathrm{IDF}$ & $\ge$80          & X          &0.9586$^{(24)}_{(25)}$      & X         & 0.9890(1)   & 0.9891(1)                            \\
$\varepsilon_\mathrm{MTF}$ & $\ge$80          & 0.9925(2)  & X                          & 0.9933(1) & X           & 0.9997                               \\
$\varepsilon_\mathrm{MCF}$ & $\ge$80          & 0.9928(2)  & X                          & 0.9935(1) & 0.9997      & X                                    \\
\hline
$\varepsilon_\mathrm{IDF}$ & 80-110           & X          &0.1315$_{(423)}^{(529)}$    & X         & 0.1325$_{(103)}^{(108)}$ & 0.1355$_{(105)}^{(111)}$\\
                    	   & 110-500          & X          &0.5500$_{(246)}^{(244)}$    & X         & 0.7188(37)               & 0.7211(37)              \\
                           & 500-7k           & X          &0.9912$_{(17)}^{(15)}$      & X         & 0.9962(1)                & 0.9963(1)               \\
                           & $\ge$7k          & X          &1.0000$_{(4)}$              & X         & 1.0000                   & 1.0000                  \\
\hline
\end{tabular}
\etab

\sfig{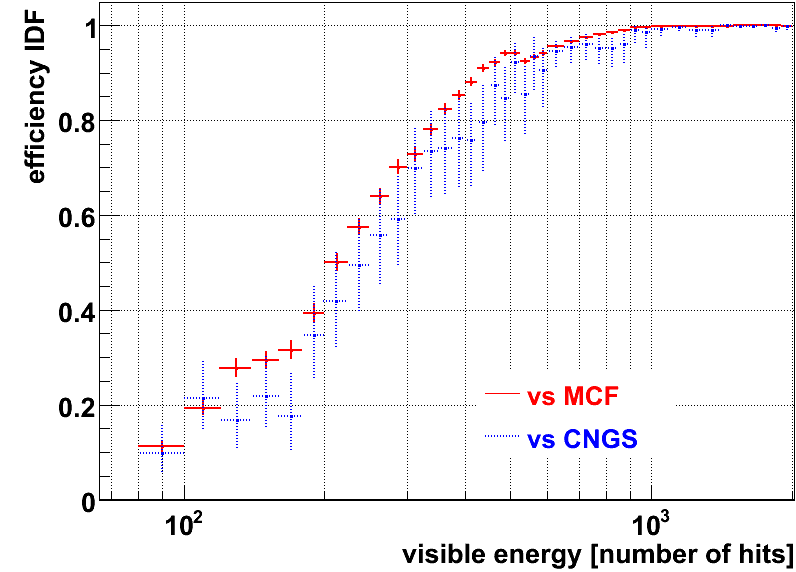}{.6}{IDF tagging efficiency against MTF-selected and against CNGS-selected events as a function of visible energy. }

\subsubsection{Mutual Efficiency Determination}
\label{sec:RelEff}

The efficiencies of the three tags can also be derived by mutual comparison. 
Each of the three flags has been tested against muon samples selected using either of the two remaining flags.
The results of the mutual tests of MTF and MCF to each other are included for completeness and should be handled with care 
as the two flags share the same set of PMTs and are therefore not truly uncorrelated. 
However, tests seeing IDF on the one side and MTF/MCF on the other are fully valid as the two detectors are completely independent. 

Impurities in the samples arise here from point-like events due to potential ``over-efficiency'' of the reference flag. 
MTF and MCF are in principle free of over-efficiencies as there is no light connection between ID and OD. 
However, small light leaks in the external steel dome increase the rate of OD\textmu-like events up to $\sim$2.5$\cdot$10$^{4}$\,cpd and this can cause accidental coincidences with point-like ID events. 
The number of such events has been statistically evaluated; we have found that for MCF (MTF) the ``over-efficiency'' above the 80 hits threshold are as low as $\sim$10$^{-5}$ ($\sim$10$^{-6}$).
The over-efficiency of IDF is not known precisely, however from the results of this section it is limited to below 0.7\%.

The resulting efficiencies are shown in section ``Mutual Efficiencies'' of \tab{TagEff}. 
The tests versus IDF imply efficiencies $\ge(99.33\pm0.01)\,\%$ for MTF and MCF. 
IDF on the other hand shows a good performance, $\ge(98.90\pm0.01)\,\%$, if integrated over the whole muon spectrum. 
However, in agreement with the results of the direct efficiency study, discrimination power decreases for low visible energies (\fig{idf}).

\subsubsection{Overall Veto Efficiency}
\label{sec:overall}

We have evaluated efficiency of the available muon tagging methods both directly and from mutual comparison obtaining consistent results that are summarized in \tab{TagEff}. 

In all Borexino analyses, muons are suppressed from the data set by the application of both OD- and ID-related flags, 
MCF being the preferred choice when available to combine with IDF. 
MTF is used in the small fraction of data when MCF is not available due to DAQ inefficiencies.
The muon veto efficiency from the application of the combined tags is at least 99.992\,\%.

\section{Muon Track Reconstruction}
\label{sec:TraRec}

The primary goal of muon tracking in Borexino is to enhance the veto of cosmogenic radioisotopes by the introduction of geometrical cuts. 
Muons produce unstable nuclei by spallation processes on their trajectory through the scintillator and the decay of such nuclei can mimic the expected signal. 
For short-lived radioisotopes, it is usually sufficient to veto all events in the wake of a muon for a duration corresponding to several half-lifes of the given isotope. 
This kind of veto is sufficient for {$^{7}$Be}-$\nu$ and geo-neutrino analyses because the cosmogenic background in these cases is due
to nuclei whose half-life is of the order of a few seconds or less. 
However, when the half-life of the dangerous nuclei is longer, the induced dead time can become very large. 
To overcome this problem, the veto of events can be applied only to a cylindrical region around the muon track instead of the whole detection volume. 
While information on the muon track was already exploited in the {$^{8}$B}-$\nu$ analyses \cite{bx10b8}, it is of uttermost importance for the CNO and pep-$\nu$ search: 
here, the most prominent background arises from cosmogenic $^{11}$C that surpasses the expected neutrino rate by an order of magnitude. 
$^{11}$C features a half-life of 20.38\,min, so a complete veto of the detector for several half-lifes following each muon would result in a dead exposure close to 100\,\%. 
Fortunately, $^{11}$C production is in 95\,\% of all cases accompanied by a knock-out neutron from $^{12}$C. 
The $^{11}$C background events can be identified by a threefold coincidence relying on spatial and time information of the muon, 
the neutron (see \sec{neu}), and the decay signal of the radionuclide as described in more detail in \cite{gal04}.

Moreover, muon tracking allows an accurate investigation of both cosmogenic neutron background 
and other muon-induced production of radioisotopes occurring in Borexino. 
Such a study is in preparation and is of particular value not only for the Borexino analyses, 
but also for other low background experiments located at LNGS or at other underground laboratories of similar depth.

\subsection{Reconstruction Algorithms}
\label{sec:RecAlg}

We have developed two independent tracking algorithms.
The {\it OD tracking} is based on the entry and exit points generated by the \v{C}erenkov cone of muons in water.
The {\it ID tracking} relies on the distinctive photon arrival time pattern created by the scintillation light front emitted by the muon track.
The results of both ID and OD tracking algorithms are later combined in a best fit {\it global tracking}.

\subsubsection{Outer Detector Tracking}
\label{sec:TraOD}

The algorithm is based on the identification of the disc-like activation profile of OD PMTs generated by ID\textmu's: 
the \v{C}erenkov cone around the track is registered by the PMTs close to the muon \textit{entry point} EP (usually on the SSS) 
and with some time delay by PMTs near the \textit{exit point} XP (either on SSS or floor of the water tank, depending on track inclination). 
High-energy cosmic muons are highly relativistic.
As the \v{C}erenkov light is propagating at the speed of light in water ($0.75c$), the muon arrives at the PMT-equipped surface first, followed by its \v{C}erenkov light cone. If the track is perpendicular to the surface, the first light is detected by the PMTs nearby the penetration point.
As light is emitted at a fixed \v{C}erenkov angle $\alpha_c = 41^\circ$, the time profile of PMT hits follows circles or (depending on the track inclination) ellipses around the penetration point, from early hits in the center to the late hits at the periphery. The number of photons per PMT decreases with the distance to the penetration point.
\Fig{cherenkov_profile} shows a view of the typical PMT activation profile.

\sfig{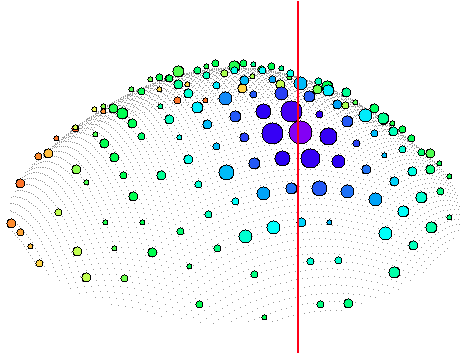}{.5}{Activation profile created by the \v{C}erenkov cone of a muon that is crossing the curved surface of the SSS: the size of the dots indicates the number of photons registered by a PMT, while colors code the photon arrival time (\textit{violet}$\rightarrow$\textit{red}).}

The tracking algorithm uses clusters of hits associated in time and space to identify EP and XP. PMT hits on the SSS and on the floor are treated separately. The spatial distance between two hits on the SSS (floor) has to be $\delta r \leq 3$\,m (4\,m) in order to affiliate the pair to the same cluster. This condition corresponds to about twice the mean distance between neighboring PMTs, which is 1.7\,m (2.2\,m). In addition, the time difference between these hits has to be smaller than 20\,ns, which is generous compared to the expected time difference between neighboring PMTs of 4.9\,ns (6.9\,ns).
In this way, separate clustering of light reflected back at the original penetration point is suppressed. 

The charge barycenters $\vec{R}_\mathrm{bc}$ of clusters are computed to serve as future entry and exit points. 
The coordinates $\vec{r}_i$ of the affiliated hits are weighted by $w_i$ considering both 
the signal charge $q_{i}$ and the PMT hit time $t_{i}$:
\begin{eqnarray} \label{eq:CluCen}
\vec{R}_\mathrm{bc} = \frac{\sum_i w_i \vec{r}_i}{\sum_i w_i} \,\,\,\,\,\,\mathrm{~with~} \,\,\, w_i(q_i,t_i) = q_i\cdot\mathrm{e}^{-(t_i-t_0)/\tau}
\end{eqnarray}
with $\tau=20$\,ns and $t_{0}$ the time of the first hit affiliated to the cluster. 
The barycenters are calculated in Euclidean coordinates, 
but radially shifted to the radius of the OD PMT photocathodes to take into account the SSS surface curvature. 
The deviation of the weighted mean is used as statistical uncertainty. 

For an ideal track, only two clusters caused by EP and XP belonging to a single track are present, making no further selection necessary. However, light reflected by the Tyvek foils creates in most cases additional fake clusters that have to be rejected. A large fraction of these clusters can be discriminated by setting a minimum for the integral number of photoelectrons as most of the light intensity is concentrated in the clusters containing direct light. The main criteria applied to identify the OD entry cluster are its total number of photoelectrons and its synchronization with the start time of the corresponding ID event. Based on the identified EP, the exit cluster is found by checking the time-of-flight difference that corresponds to the trajectory of the muon traveling at speed of light in vacuum. 
Moreover, a reconstructed track is only considered valid if it crossed the sphere, as due to the synchronous trigger of ID and OD only ID\textmu's are considered.

There are events where only the EP can be identified: if the muon stops inside the ID, there is no XP. 
Alternatively, if the track leaves the ID through the optically inactive lower hemisphere at a flat angle, it produces only a small amount of light that is sometimes too faint to be detected. 
In these cases, it is vital to avoid the identification of a reflection light cluster as XP.

\subsubsection{Inner Detector Tracking}
\label{sec:TraID}

The ID muon reconstruction is based on several successive fits to the distinct photon arrival time patterns generated by the tracks. The muon \textit{entry point} EP is found by fitting the arrival time distribution of the first hits detected inside the ID. We then introduce the plane $P_{exo}$, defined by the muon entry and exit points as well as the SSS origin. $P_{exo}$ is a symmetry plane that divides the photon arrival pattern of the PMTs in two identical halves. It is identified by several fits to the arrival time patterns along rings of PMTs in rising distance from EP. As the muon track lies on $P_{exo}$, both EP and the \textit{exit point} XP are in this plane (\fig{nid_geometry}). Finally, a fit to the time distribution of the PMTs in $P_{exo}$ is used to find the discontinuity in the time pattern around XP. This local minimum at XP is generated by the tip of the muon light cone penetrating the sphere surface.

\sfig{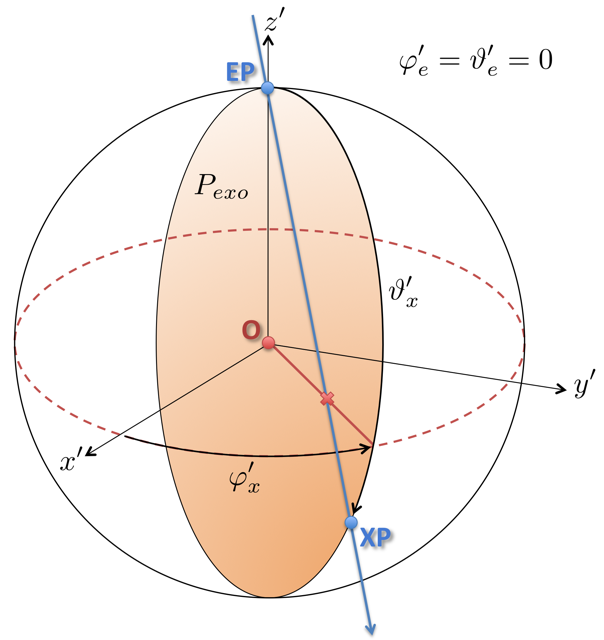}{.5}{Geometry defining the muon track: the coordinate system was rotated to shift the entry point EP to the zenith of the SSS. The symmetry plane $P_{exo}$ of the photon arrival time distribution is defined by the sphere origin O, EP and the exit point XP. The azimuthal coordinate $\varphi_x'$ of XP is determined by fits to the arrival time distributions on horizontal circles of PMTs (parallel to the red dashed circle). The zenith angle $\vartheta_x'$ is found by a fit to the arrival times along the intersection of $P_{exo}$ and the SSS surface (see \fig{muon_entry_fits}).} 

In fact, the light front generated by a muon in the ID is described by a superposition of scintillation and \v{C}erenkov light. 
While scintillation light is clearly dominant for the part of a track located in the IV, the DMP added to the buffer suppresses the scintillation light output of the PC \cite{bib:DMP}. 
For B\textmu's and for the buffer portion of the track of IV\textmu's, light is generated in comparable amounts by scintillation and the \v{C}erenkov effect. 
However, the resulting shape of the light front will be similar in IV and buffer: 
as depicted in \fig{IDlightfront}, the superposition of spherical waves of scintillation light along the muon track forms a conical light front centered around the muon trajectory. 
The opening angle is identical to the \v{C}erenkov angle as it depends merely on the light speed in the scintillator medium. In addition to the cone, a backward-running spherical light front is caused by the scintillation light only. 

\dfigp{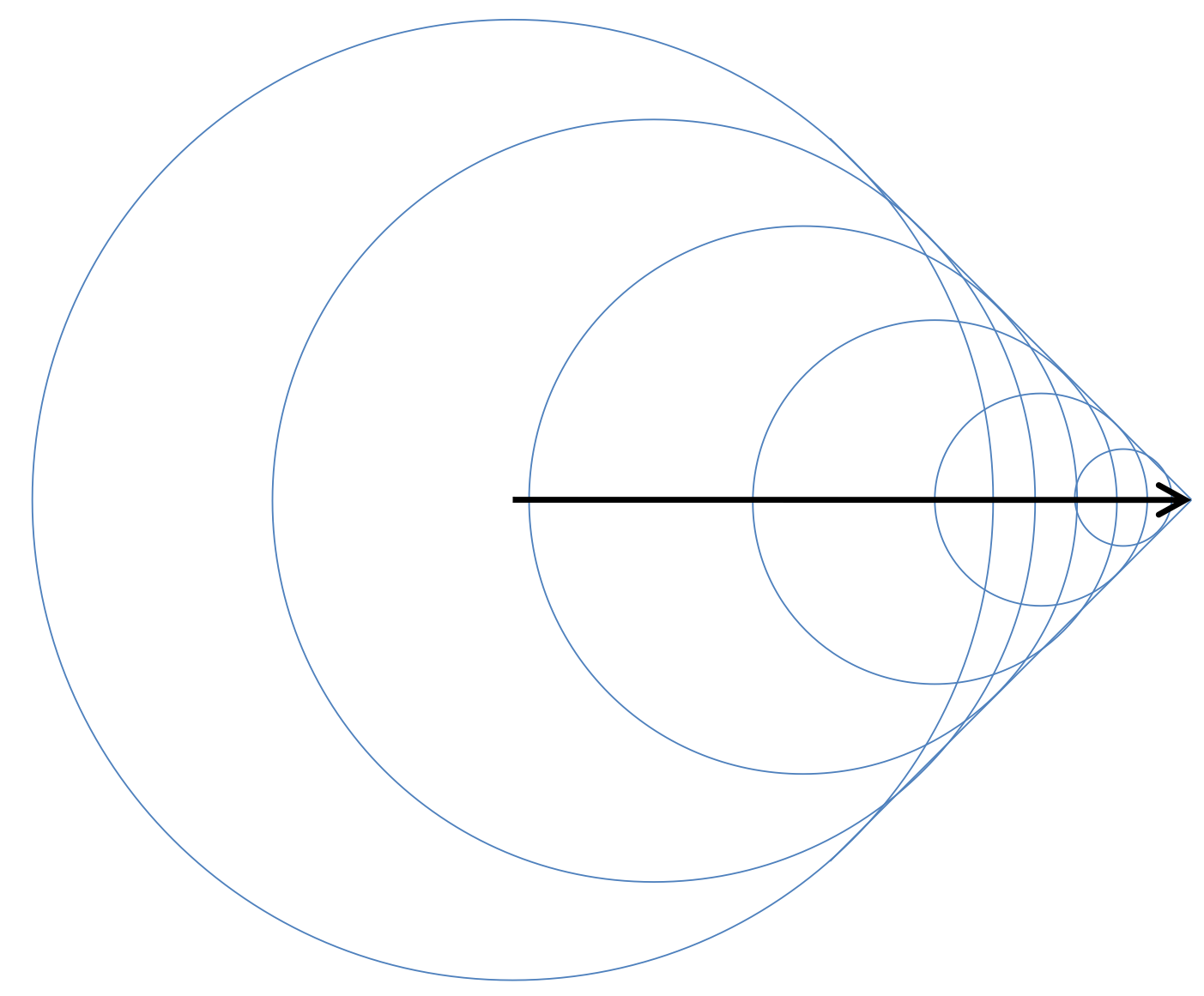}{.43}{A muon traversing the ID  creates light by scintillation processes. The superposition of spherical waves along the track creates a light front that is in forward direction identical to a regular \v{C}erenkov cone. In addition, a spherical light front propagating backwards is created. }
          {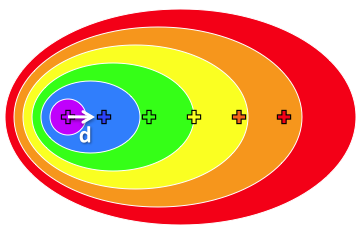}{.55}{A muon entering the ID creates ellipses of isochronous PMT hits. The entry point is in one focus. The very first hits (symbolized by the violet area) should be almost centered on the entry point position. The relative offset of the barycenters for later isochronous hits bears information on the inclination vector of the muon track $\vec d$. It is used in an alternative algorithm described in \cite{wur09phd}.}

\subsubsection*{Entry Point Reconstruction}

The backward-running light reaches the SSS first at the muon entry point EP. 
For a first approximation of the EP position, the ID tracking analyzes the time pattern of the hits detected in the first 5\,ns of the muon event, corresponding to the first 70\,cm of the track inside the ID:  similar to the OD, the quenched scintillation light of the buffer produces ellipses of isochronous PMT hits for which EP lies approximately in a focus (\fig{IDTimeDist}). 
As the charge information of ID PMTs is not reliable due to the large muon light output, only timing is used to compute the barycenter according to \eq{CluCen}.

Based on the first guess, the time distribution of hits around EP is fitted with a function $t_e(\phi)$ optimized to find the minimum arrival time that corresponds to EP: 
for this, the first-approximation barycenter is shifted by rotation to the equatorial plane (azimuth angle $\varphi=0$, zenith angle $\vartheta=\pi/2$) in order to avoid the topological effects of the spherical coordinates that arise mainly on the poles. The empirical fit function 
\beq
\label{eq:EntFit}
t_e(\varphi') = a^0 + \left\{ \begin{array}{ll}
a^-\cdot\sin(c^-(\varphi_e'-\varphi')) & \textrm{if $(\varphi'<\varphi_e')$}\\ a^+\cdot\sin(c^+(\varphi'-\varphi_e')) & \textrm{else}
\end{array} \right. 
\eeq
is of V-like shape. 
It allows for different curvatures $c^\pm$ and slopes $a^\pm$ of the arms in order to accommodate the curved surface geometry and a possible relative inclination of the track that results in differences in the light propagation time. 
Also a small time offset $a^0$ is included. 
\Fig{muon_entry_fits} shows the time distribution of hits dependent on the angle relative to the barycenter. 
Plots a) and b) correspond to two perpendicular orientations along the sphere surface that correspond to the $\varphi'$ and $\vartheta'$ planes in the transformed coordinates.
The deviations $\varphi_e'$ (and $\vartheta_e'$) from $\varphi'=0$ ($\vartheta'=0$) that can be extracted from the fits are transformed back to regular Borexino coordinates to obtain the corrected position of EP.  

\sfig{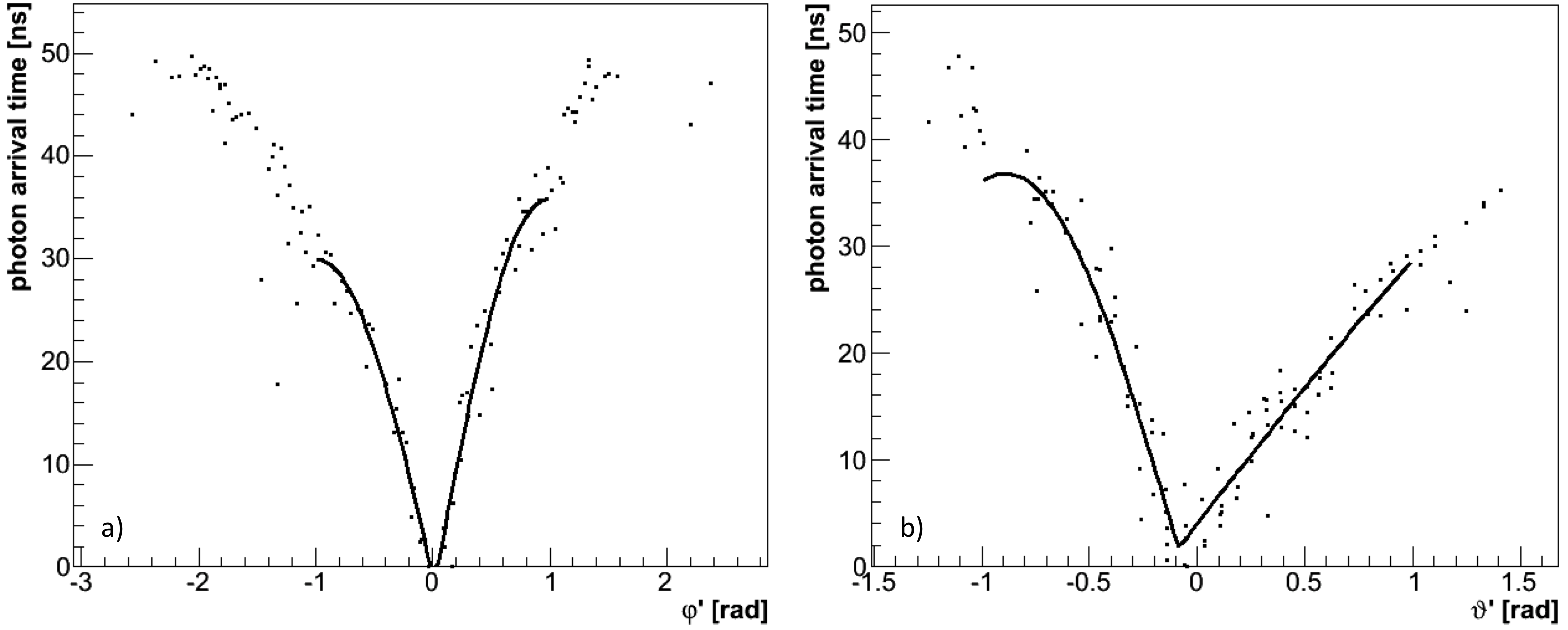}{1}{The arrival time pattern at the entry point registered by ID PMTs as a function of zenith $\vartheta'$ (a) and azimuth $\varphi'$ (b) angles corresponding to PMT positions on a geodetic circle around the SSS. The two circles intersect at the supposed positions of the entry point at $\varphi'=\vartheta'=0$.}

\subsubsection*{Finding the Symmetry Plane}

To identify the muon exit point XP, the algorithm searches for the symmetry plane $P_{exo}$ defined above. To find it, the PMT coordinates are again rotated, this time to shift EP to the zenith. The corresponding geometry is shown in \fig{nid_geometry}. 
If in these coordinates a horizontal plane of PMTs is chosen and the arrival time is plotted as a function of the azimuth of the transformed PMT positions, 
the observed time pattern is again of V-like shape. The minimum is located at the angle $\varphi_x'$ corresponding to $P_{exo}$ in the transformed coordinates. The fit function
\begin{eqnarray}
t_x(\varphi') = c_{pc}^{-1} \cdot \sqrt{ R_\mathrm{S}^2+R_\mathrm{I}^2-2R_\mathrm{S}R_\mathrm{I} (\cos\varphi'\cos\varphi_x'+\sin\varphi'\sin\varphi_x') }+t_0
\end{eqnarray}

\noindent reflects the symmetric dependence of the arrival times on $\varphi'$. It is inspired by the time distribution expected if the track crossed the chosen plane of PMTs exactly perpendicularly: in this case, $R_\mathrm{S}$ and $R_\mathrm{I}$ would denote the radius of the sphere and the impact parameter of the muon track relative to the sphere center, respectively, while $c_{pc}$ denotes the light speed in liquid scintillator. 

The algorithm performs this fit on twelve horizontal planes of PMTs in rising distance below EP, each returning a best-fit value for $\varphi_x'$. 
Finally, the mean of these results is used, weighting each result with the fit uncertainty. 
In principle, this algorithm returns a very precise result for $\varphi_x'$ as the information of almost all of the PMTs is taken into account.
However, if EP position is off by a certain angle relative to $P_{exo}$, 
this will cause a sine-like distortion of the $\varphi_x'$ fit results (see below). 

\subsubsection*{Exit Point Identification}

To find the second coordinate angle $\vartheta_x'$, the PMT circle defined by $P_{exo}$ is selected.
The corresponding arrival times are drawn in \fig{muon_exit_fits}a as a function of the (transformed) zenith angle $\vartheta'$. Here, the EP is located at $\vartheta_e'=0$. 
The arrival time distribution shows 
a discontinuity in the rise 
for $\vartheta'\in[0,\pi]$. 
The kink is caused by the projection of the \v{C}erenkov-like light cone on the surface defined by the SSS, the local minimum corresponding to $\vartheta_x'$. 
The fit function 
\beq
t_x(\vartheta') = \frac{R_\mathrm{S}\cdot\sqrt{2-2\cos\vartheta'}}{c\sqrt{1 - \frac{1}{n^2}}\cdot \left( \sin(\arccos(\frac{1}{n}) - \frac{1}{2}|\vartheta'-\vartheta_x'|)  + n\sin( \frac{1}{2}|\vartheta'-\vartheta_x'|) \right)} + t_0
\eeq 
is derived from an analytical time of flight calculation based on the sphere radius $R_\mathrm{S}$, the speed of light $c$ and the refractive index $n$. The function assumes light emission from the muon track under an angle corresponding to the \v{C}erenkov condition and provides an excellent fit. However, the function is only valid close to the true value of $\vartheta_x'$; for other regions, the observed arrival times are changed due to the geometrical cut-off caused by the sphere. Therefore, the fit routine is performed only on a limited interval surrounding a first prediction of $\vartheta_x'$ based on the visible light output: 
\fig{ci_vs_dnhits} shows the basic relation between the number of hits registered in the ID and the track impact parameter $R_\mathrm{I}$ determined with the OD tracking algorithm. The fit routine is iterated several times, each time starting from the $\vartheta_x'$ computed in the previous step and progressively narrowing the fit range at each step.

This method works best above 4000 hits, which corresponds to tracks with an impact parameter $R_\mathrm{I}$ of less than 4.5\,m (\fig{ci_vs_dnhits}). In the region from 2500 to 4000 hits ($R_\mathrm{I}\in [$4.5\,m;\,6\,m]), the scatter in the arrival times becomes larger so that the distribution has to be averaged over angular bins of $\Delta\vartheta=\pi/30$ in width. In addition, a slightly modified fit function that stresses the discontinuity around XP has to be applied in order to obtain reliable results. Below 2500 hits, $\vartheta_e'$ and $\vartheta_x'$ lie too close to be discerned in the fit, and $\vartheta_x'$ collapses to 0. In this case, only EP coordinates are later on used in the global track (see below).

\sfig{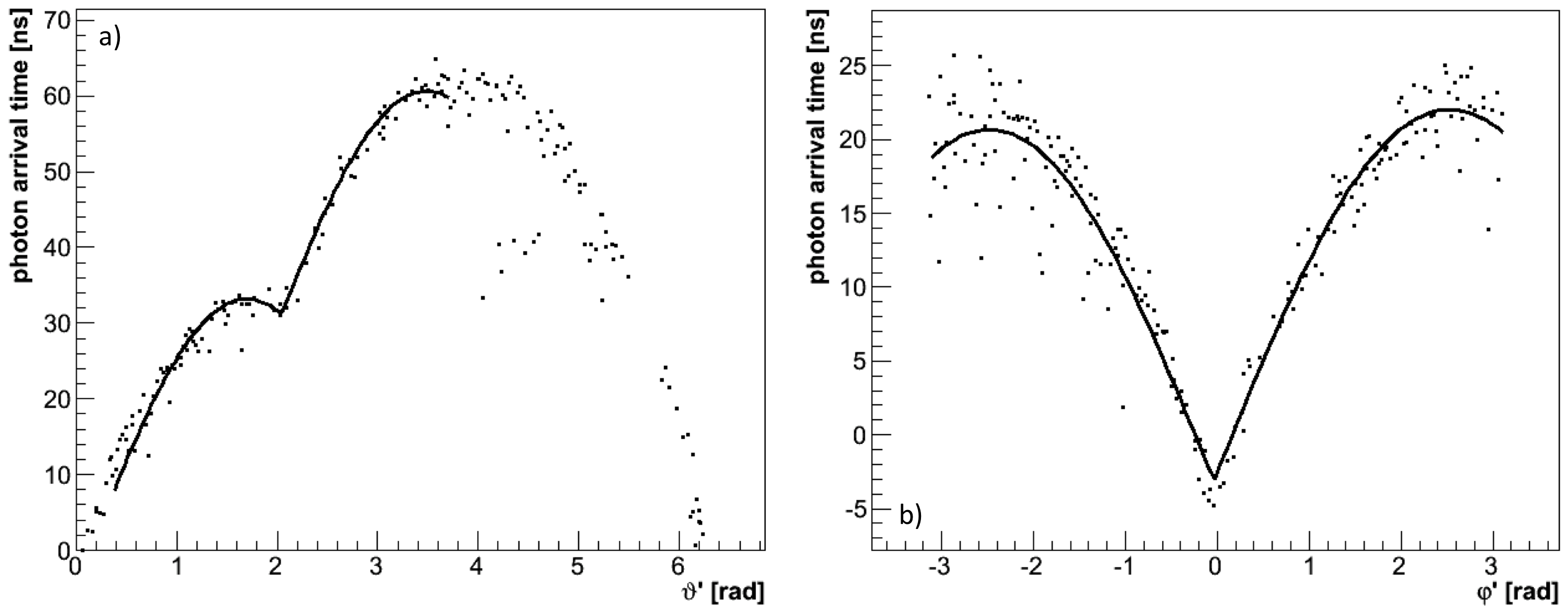}{1}{Photon arrival time distribution around the exit point XP: a) distribution $\vartheta'$ for PMTs on the symmetry plane 
of the sphere $P_{exo}$ (see \fig{nid_geometry}); 
b) distribution $\varphi'$ for PMTs corresponding to a geodetic circle intersecting at $\vartheta_x'$.}

\sfig{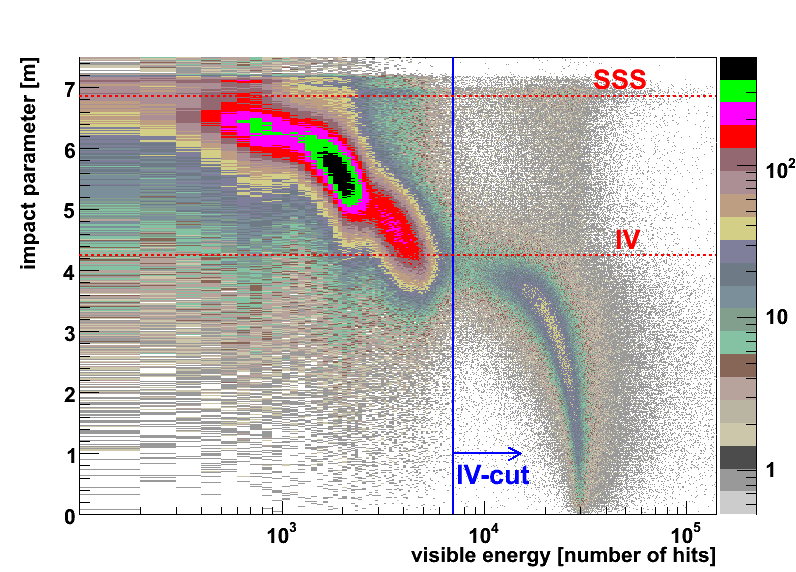}{.6}{The impact parameter determined by OD tracking versus the number of ID hits in
the muon event. Two main bands are visible in the regions below and above 7000 hits: they are caused by
B\textmu's and IV\textmu's respectively. The IV band is strongly bent due to the relation of path length and of the impact parameter.}

Finally, the fit on $t_x(\varphi')$ is repeated based on the effective function described by \eq{EntFit}. The result is shown in \fig{muon_exit_fits}b. This last step is used to apply a small correction to $\varphi_x'$ in cases in which the symmetry plane $P_{exo}$ determined before is inclined relative to the true muon track.
XP coordinates are then obtained by transforming back $(\varphi_x',\vartheta_x')$ to the original reference frame.

\subsubsection{Global Tracking}
\label{sec:TraGlo}

Combining the information provided by OD and ID muon tracking, a global track can be defined representing the best fit to the (up to four) data points available for a single track. 
Different from the other tracks, the information on the global track is not saved in the form of points on the track but using a set of four fit parameters. 
The three dimensional track $\vec T(x)$ is parameterized as~:
\begin{eqnarray} \label{eq:GloTra}
\vec T(x) = \left(\begin{array}{c}0 \\ \alpha \\ \gamma \end{array}\right) + x \left(\begin{array}{c}1 \\ \beta \\ \delta \end{array}\right).
\end{eqnarray}
The reduced notation is $y(x) = \alpha+\beta x$, $z(x) = \gamma + \delta x$. The latter equations already indicate that the 3D fit can be split into two independent 2D fits to projections of the track on the $xy$ and $xz$ coordinate planes. To preserve the information on the direction of motion of the muon, a fifth variable indicating the track orientation is necessary. \Fig{GLtracks} shows the resulting track in a graphical representation. The four track points including their uncertainties (indicated by boxes) are shown along with the fitted track; in addition, time and charge profiles provided by ID and OD are included.

\sfig{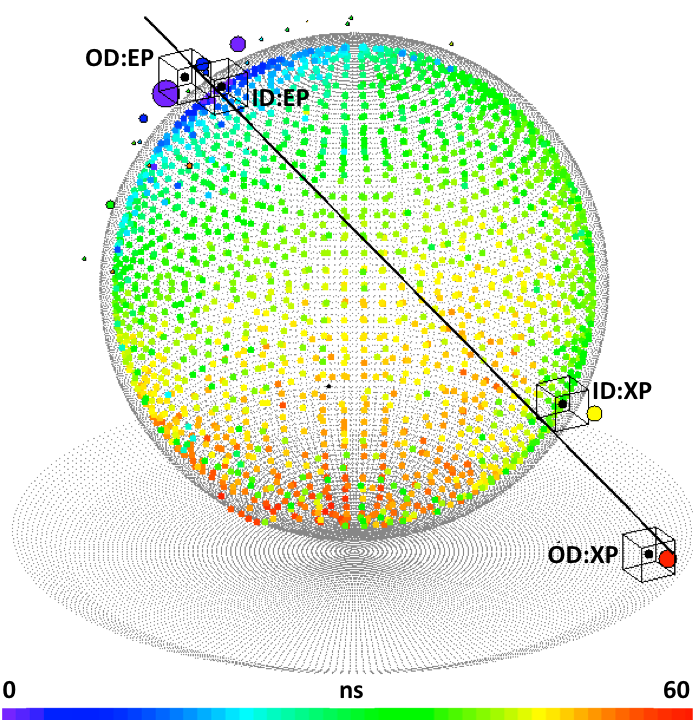}{.5}{Graphical description of the global muon tracking. The track points provided by ID and OD reconstruction algorithms are shown, the boxes indicating their uncertainties. Also included is the information on the ID PMT hit time distribution (\textit{small colored dots}) and the OD hits (\textit{colored circles}).}

There are a number of cases in which 3 out of 4 points are reasonably aligned while a fourth point seems to be misreconstructed. 
If the reduced $\chi^2$ value is $\chi^2/\mathrm{ndf} > 3$, four additional fits are performed that each omit one of the track points and the fit with the lowest $\chi^2/\mathrm{ndf}$ is chosen.

\subsection{Tracking Performance}
\label{sec:perform}

The parameters entering the tracking algorithms were tuned using a MC simulation of muon events. 
The evaluation of the accuracy of the tracking instead has been performed on real data. Several aspects are evaluated:
~\\~\\
\noindent The \textit{efficiency} of the tracking can be tested by the ratio of cosmic muon events that feature reconstructed tracks: complete tracks with both entry and exit point are found for 98.5\,\% of all muon events for the OD tracking. The efficiency of the ID tracking is lower, 87.9\,\%. The reduction is mostly due to the minimum energy requirement of 2500 hits; above this threshold, the efficiency is 96.9\,\%. The global tracking algorithm returns a fitted track in 96.4\,\% of all muon events, corresponding to 99.3\,\% of the cases in which at least two points were found by ID or OD tracking. These values on their own have no further implication concerning the quality of the reconstruction. 
~\\~\\
\noindent The \textit{angular resolution} of the tracking can be determined by utilizing CNGS muon events, generated in the rock upstream of the Borexino detector and crossing both Borexino and OPERA \cite{aga09, aga10, acq09}. 
A track-by-track comparison between the two experiments is possible, returning resolutions of 3$^\circ$ for OD/global tracking and 5$^\circ$ for ID tracking. 
In addition, a high-resolution External Muon Tracker (EMT) has been mounted on top of the Borexino steel dome in late 2009 (see \fig{muon_definitions}), providing a small sample of well-determined muon tracks crossing Borexino. 
Statistics of this analysis is still limited, but the resulting resolution is comparable to the one obtained from CNGS muons. 
The results of CNGS and EMT analyses are complementary as they test horizontal and vertical muon tracks, respectively. 
~\\~\\
\noindent The \textit{lateral resolution} (perpendicular to the muon track) can be derived from a track-by-track comparison with OPERA, returning a resolution of 40-50\,cm (35\,cm for IV\textmu's). 
Statistics for EMT are in this case too low to return meaningful results. 
An alternative is the analysis of the distribution of cosmogenic neutrons relative to their parent muon track. 
Here, a resolution of $\sim$35\,cm for IV\textmu's is obtained.
~\\~\\
\noindent In the following, CNGS, EMT and cosmic results are discussed in more detail. 
While the performances of the individual tracking algorithms depend in detail on the applied test, consistent values can be found. 
Generally speaking, vertical and through-going muon tracks are best reconstructed by the OD and global trackings, 
while horizontal and (semi-)contained tracks are reproduced best by the ID algorithm.
 
\subsubsection{CNGS Muons}
\label{sec:perform_cngs}

As described in \app{cngs}, the CNGS beam arrives at the LNGS at an inclination angle of $\theta=93.2^\circ$. Relative to CNGS, CERN is located at an astronomical azimuth of $\varphi=305^\circ$, almost exactly coinciding with the orientation of Hall C. \Fig{CngsDis} shows the azimuth and inclination angle distributions obtained by Borexino (and OPERA) tracking.

\dfigsr{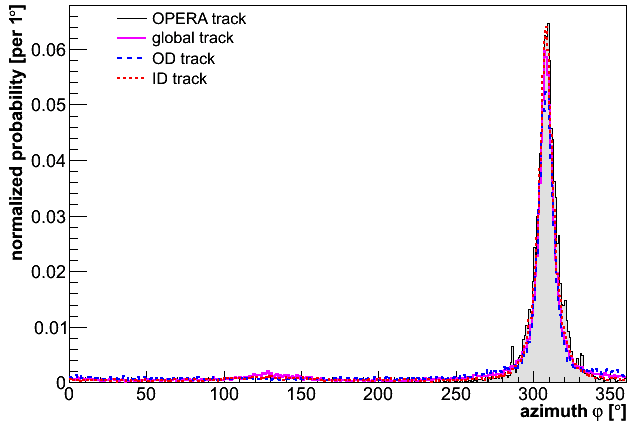}{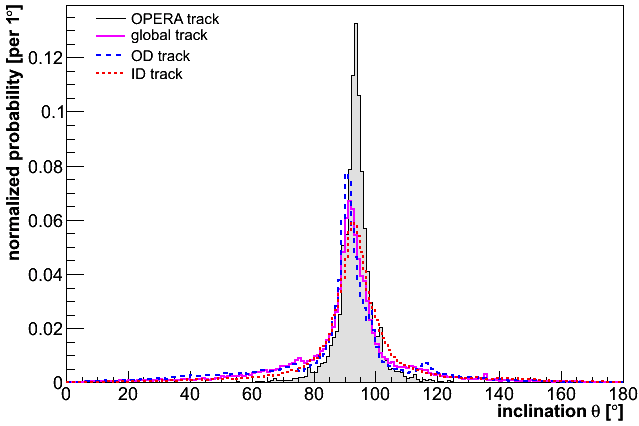}{CngsDis}{Angular distributions of muons produced by the CNGS $\nu_{\text{\textmu}}$ beam. 
Tracks in Borexino are required to cross both ID and OD. The distributions reconstructed by OPERA are shown for comparison.}

A test of the tracking angular resolution is possible by an evaluation of the CNGS events that were reconstructed both in Borexino and in OPERA. 
The angular resolution for muon tracks in OPERA is of the order of mrad and by far superior to the performance of the Borexino tracking \cite{zim06}. 
Therefore, the intermediate angle $\alpha$ between the reconstructed track orientation vectors of both experiments corresponds almost exactly to the angular resolution of the Borexino tracks:
\beq
\label{eq:alpha}
\cos\alpha  = \hat\mu(\varphi,\theta)\cdot\hat\nu(\phi,\vartheta) = \sin\theta\sin\vartheta(\cos\varphi\cos\phi+\sin\varphi\sin\phi) +  \cos\theta\cos\vartheta
\eeq
where $\hat\mu(\varphi,\theta)$ and $\hat\nu(\phi,\vartheta)$ represent the track orientation vectors.

In the period from 2008 to 2010, 2797 events featuring tracks in both  Borexino and OPERA detectors are found ($\sim$5\,\% of the overall number of Borexino ID on-time events). 
Several selection criteria are applied to allow a comparison: in Borexino, events are required to produce light synchronously in ID and OD, in order to obtain a sample of through-going muon tracks, generated in the rock upstream of the detector. 
In OPERA, only muon tracks featuring a momentum greater than 20\,GeV are selected. 
This reduces the effect of muon forward-scattering that might otherwise bend some of the muon tracks on their passage from Borexino and OPERA. 
The resulting samples of coincident muons feature a few hundred events.

\Fig{cngs_alpha} shows the distribution of $\alpha$ as defined in \eq{alpha} for the comparison of OD and OPERA tracking. 
Each entry is weighted with $1/\sin\alpha$ to take the spherical geometry into account. 
The distribution can be fitted by the sum of a Gaussian centered on $\alpha=0$ and a constant term that accounts for misaligned tracks for larger values of $\alpha$. 
The latter are mostly caused by muons emerging from $\nu_{\text{\textmu}}$ charged-current interactions inside the Borexino ID and missing a reliable entry point for both ID and OD tracking.

\Tab{TraCngs} lists the fit results for the Borexino tracking algorithms. The angular resolutions determined in this way range from 3$^\circ$ (OD, global) to 5$^\circ$ (ID tracking). If only IV\textmu's are selected, the resolution improves to $\sim$2.5$^\circ$.

\dfigpr{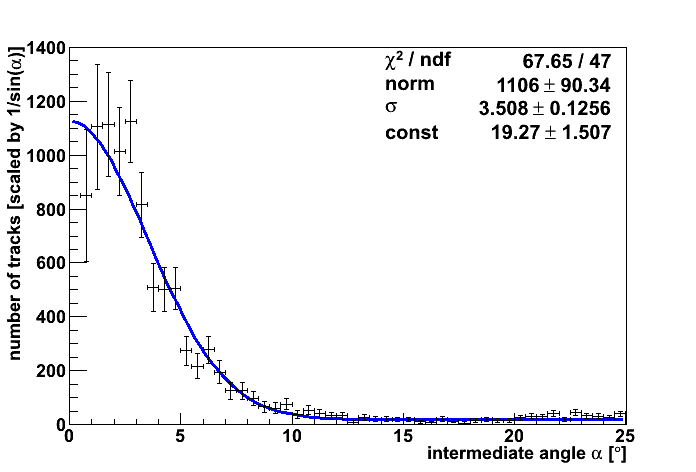}{Intermediate angle $\alpha$ between the direction vectors of muon tracks reconstructed in OPERA and Borexino OD tracking. The width of the Gaussian fit corresponds to the OD angular resolution.}
	  {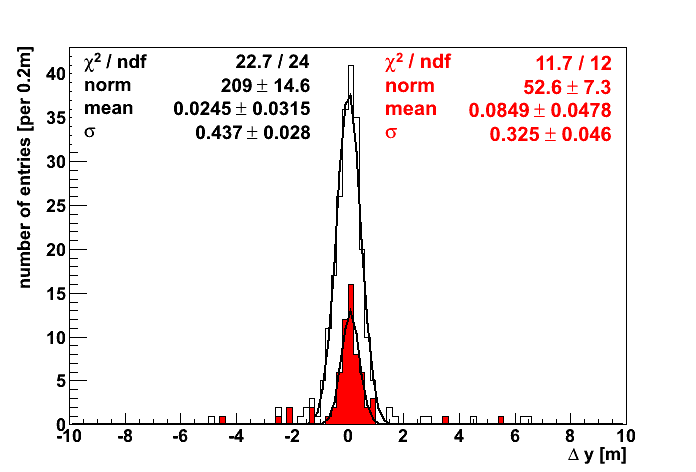}{y-distance distribution between reconstructed tracks in the Borexino yz-plane, comparing OPERA and Borexino global tracking. The lateral resolution $\sigma_y$ improves if only IV\textmu's are selected (\textit{shaded red area}).}

As not only the track direction but also the absolute track coordinates are known for both Borexino and OPERA, a determination of the lateral resolution is possible. For this, the distribution of relative distances of the track penetration points in the Borexino $yz$-plane (perpendicular to the CNGS beam direction) can be used. \Fig{cngs_global_lateral} shows the $y$-distance distribution between Borexino global and OPERA tracks. The center was artificially set to 0, as the relative coordinates of the two detectors are known only at an accuracy of 1\,m. However, the distance distribution can be approximated by a Gaussian, the width $\sigma_y$ corresponding to the lateral resolution. For the vertical $z$ direction, the same fit can be applied, returning $\sigma_z$.

The resolution values for all trackings are reported in \tab{TraCngs}. 
Considering all ID\textmu's, there is no difference in $\sigma_y$ and $\sigma_z$ for ID tracking, 
but large values for $\sigma_z$ are observed for the global and especially for OD tracking. 
This effect is caused by the lower non-instrumented section of the SSS outer surface -- 
horizontal OD tracks passing this section will be reconstructed either too close to the equator or to the OD floor. Obviously, also the quality of the global tracking is affected. However, the effect vanishes if only tracks passing the PMT-equipped upper part of the SSS are selected. If only IV\textmu's are selected (more than 7000 hits in Borexino), the lateral resolution improves in both $y$ and $z$ to $\sim$35\,cm for all trackings, $\sigma_y\approx\sigma_z\approx\sigma_\mathrm{lat}$.

\btab
\captab{TraCngs}{Fit results for angular and lateral resolution obtained from CNGS muon tracks observed by Borexino and OPERA.}
\begin{tabular}{|l|ccc|}
\hline
ID\textmu's & $\sigma_\alpha$\,[$^\circ$] & $\sigma_y$\,[cm] & $\sigma_z$\,[cm] \\
\hline
OD tracking		& 3.51$\pm$0.13 & 53$\pm$4 & 198$\pm$13 \\
ID tracking		& 5.13$\pm$0.25 & 42$\pm$3 & 40$\pm$5 \\
GL tracking		& 2.83$\pm$0.13 & 44$\pm$3 & 87$\pm$12 \\
\hline
\hline
IV\textmu's & $\sigma_\alpha$\,[$^\circ$] & $\sigma_y$\,[cm] & $\sigma_z$\,[cm] \\
\hline
OD tracking		& 3.01$\pm$0.15 & 28$\pm$7 & 45$\pm$7 \\
ID tracking		& 2.44$\pm$0.19 & 36$\pm$5 & 31$\pm$6 \\
global tracking & 2.31$\pm$0.13 & 33$\pm$5 & 35$\pm$7 \\
\hline
\end{tabular}
\etab

\subsubsection{External Muon Tracker}
\label{sec:perform_emt}

The external muon tracker (EMT) is a support system allowing the determination of the coordinates 
of a subset of muon tracks at high precision in order to test the Borexino reconstruction algorithms. 
Originally developed as a prototype of the OPERA muon tracker, the EMT is based on layers of drift tubes. 
Whereas in the original setup only a 2D track reconstruction was possible, 
the tracker has been modified for a three dimensional tracking by crossing the modules. 
It has been mounted on a platform above the Borexino steel dome. 
The location was chosen 4.6\,m off the vertical symmetry axis of the detector and above the dome 
in order to sample muons crossing all sub-volumes of the detector (\fig{muon_definitions}). 

\sfig{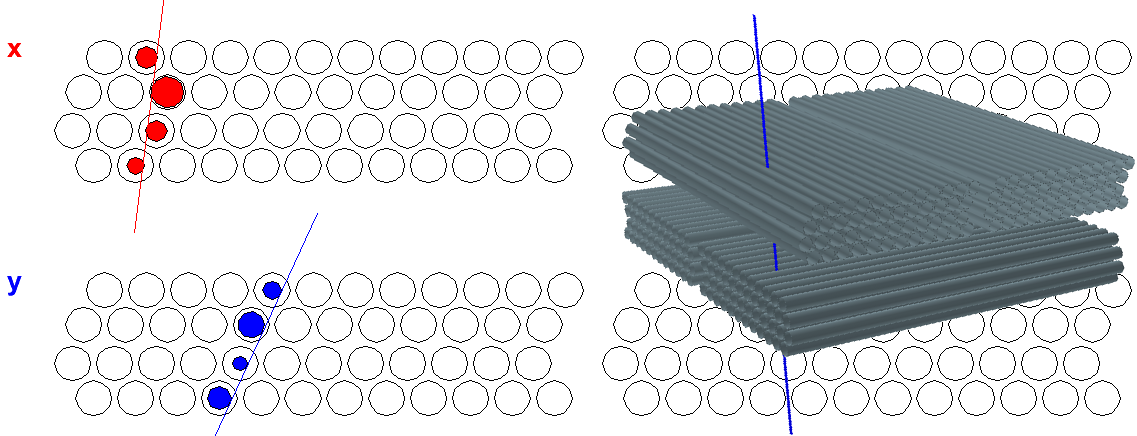}{.8}{Sample event from the EMT track display. Open circles correspond to a head-on view of the drift tubes in upper (x) and lower (y) layer. The filled colored circles indicate the observed drift times, from which the track projections in the yz- and xz-plane are derived. The overlay shows a schematic three-dimensional view of the spatial arrangement of the EMT drift tubes.}

As depicted in \fig{cmt}, the EMT modules are arranged in two arrays of drift tubes that are oriented perpendicular to each other, 
one above the other with an interspace of a few centimeters caused by the endplates of the modules. 
Each array contains 96 closely packed drift tubes stacked in four layers of 24 tubes. 
Each tube is 1.05\,m in length and 3.8\,cm in diameter, and is equipped with a gold plated tungsten wire of 45\,\textmu m diameter in the center as an anode. 
The total area covered is close to 1\,m$^{2}$. 
This corresponds to an expected muon rate of $\sim$25 events per day. 
Considering the position of the EMT with respect to Borexino, the cosmic muon flux 
and angular distribution in hall C, we estimate the daily number of muons crossing both detectors: An average of 9.6 muons/day is expected to cross both the EMT and the OD, of which 6.1 also cross the ID and 2.5 the IV.

The drift tubes are operated with an 80:20-mixture of argon and carbon dioxide as a drift gas. Spatial reconstruction relies on the drift times of electrons caused by the muon through ionization of the drift gas. The drift time in a hit tube depends on the radial distance of the muon track from the axial wire. Each array of tubes provides 2D information on the muon track. Combined, the system provides a three-dimensional track.

The signals of all tubes are read out by a customized TDC board which is also used in OPERA, connected to a PC close to the EMT. 
Data are written to its local hard drive. 
The trigger is formed by two layers of plastic scintillator panels of 2\,cm thickness that are read out by 8 PMTs. 
The trigger is issued in case of a coincidence signal in both planes.
The trigger logic is based on NIM modules.
Low and high voltages are also generated on-site in separate modules. 
A copy of the logic trigger signal is sent to the Borexino main DAQ via a 30\,m long coaxial cable. 
This allows us to relate corresponding events of the EMT and Borexino by time synchronization.

Laboratory tests give an angular resolution better than 5\,mrad and 
a spatial resolution of $\sim$300\,\textmu m, by far surpassing the resolution 
achieved for the Borexino tracking algorithms. 
The position relative to the detector center was determined with an accuracy of $\sim$1\,cm. 
The EMT was installed in late 2009, and has been taking data since spring 2010. 
However, due to the relatively low efficiency of the current trigger system the available statistics is rather low.
A replacement of the trigger system with a higher efficiency one is foreseen in early 2011.

As in the case of CNGS muons, the angular resolution $\sigma_\alpha$ of the tracking algorithms 
is determined by a track-by-track comparison, relying on the intermediate angle $\alpha$ of \eq{alpha}). 
A typical histogram for the ID tracking is shown in \fig{emt_id_alpha}, the fit results are listed \tab{TraEmt}. 
In spite of the low statistics, the obtained values for $\sigma_\alpha$ are comparable to the CNGS results presented above.

\bfig
\begin{minipage}[u]{0.59\textwidth}
\getfigw{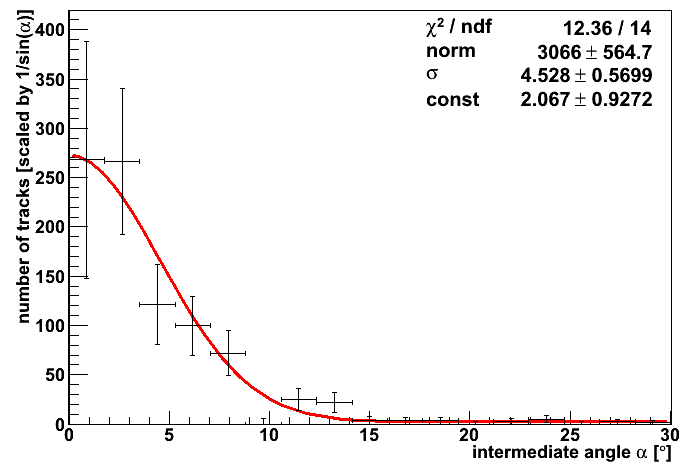}{1}
\capfig{emt_id_alpha}{Intermediate angle $\alpha$ between the direction vectors of muon tracks reconstructed by the EMT and Borexino ID tracking. The width of the Gaussian fit corresponds to the ID angular resolution.} 
\end{minipage}
\hfill
\begin{minipage}[u]{0.39\textwidth}
\centering
\captab{TraEmt}{Angular resolution as obtained by the comparison of Borexino and EMT trackings. $N$ is the number of tracks available for comparison.}
\begin{tabular}{|l|cc|}
\hline
& $\sigma_\alpha$ [$^\circ$] & $N$ \\
\hline
OD tracking & 3.9$\pm$0.3 & 125\\
ID tracking & 4.5$\pm$0.6 & 82\\
GL tracking & 3.4$\pm$0.1 & 125\\
\hline
\end{tabular}

\end{minipage}
\efig

\subsubsection{Cosmic Muons}
\label{sec:perfom_cosmic}

Cosmic muons are by far the largest muon sample available in Borexino. 
They allow a qualitative test of the angular resolution of the tracking by comparison 
of the obtained angular distribution to the earlier results of the MACRO detector \cite{macro95}. 
Moreover, cosmic muon tracks reconstructed as upward-going make it possible to determine the fraction of misreconstructed events. 

The topology of the Gran Sasso mountain massive is imprinted on the cosmic muon flux detected at LNGS. 
The underground halls were dug right under the summit of Monte Aquila, 
which is part of a small mountain ridge on the northern verge of Campo Imperatore, extending from West to East. 
Unlike in deep underground mines in which most of the muons arrive from straight above ($\cos\theta$=1, $\theta$ being the zenith angle), 
the reduced shielding towards the mountain flanks causes additional contributions at lower values of $\cos\theta$. 
The mountain profile (see app. A of \cite{macro95}) reflects in a complicated angular distribution.
Fig.\,\ref{fig:PhiDis} shows the measured distributions for azimuth $\varphi$ and inclination angle $\theta$  for cosmic muon tracks detected by Borexino. 
For comparison, the angular distributions measured by the MACRO experiment \cite{macro95} are indicated as grey-shaded areas. 

Only muon tracks reconstructed at $\cos\theta>0.5$ were selected for the shown azimuth distribution. This corresponds to the limited inclination angle acceptance of MACRO and eases the comparison. 
For both MACRO and Borexino the tracks peak around $\varphi$=30$^\circ$ (N) and $\varphi$=170$^\circ$ (S). The $\cos\theta$ distribution shows a widening in the Borexino trackings compared to MACRO. This indicates an inferior track reconstruction in Borexino, OD and global tracking performing slightly better than ID tracking. 

\dfigsr{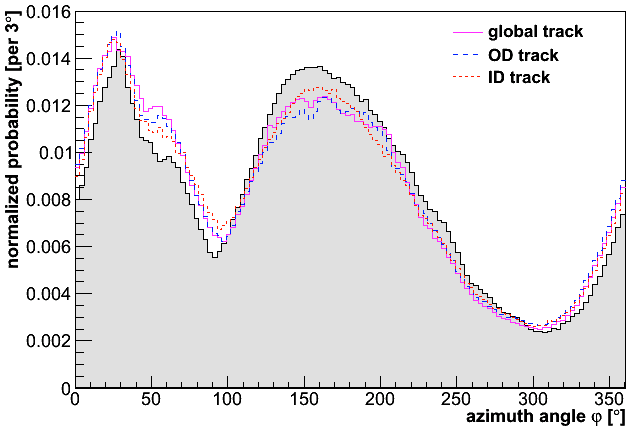}{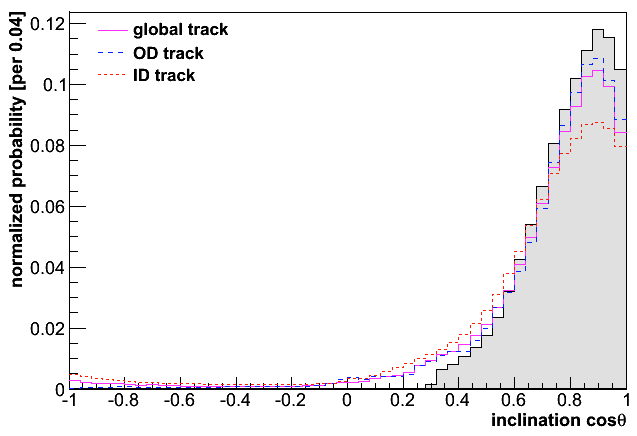}{PhiDis}{ Azimuth $\varphi$  and inclination $\theta$ distributions for cosmic muon tracks detected at the LNGS. The colored curves indicate the result of the three Borexino tracking modules.  
The result of  MACRO \cite{macro95} is the grey shaded area. }

All three tracking algorithms show a low but relatively constant number of tracks at $\cos\theta$$<$0. Upward-going muons are supposed to be very rare as they are solely produced by antipodal atmospheric neutrinos. The tracks at $\cos\theta$$<$0 are far too numerous and must be regarded to a large extent to be mis-reconstructed. \Tab{TraSig} shows the fraction of upward-going tracks in dependence of the visible energy. Overall, about 2-5\,\% of tracks are reconstructed as rising. They are mostly caused by a misalignment of EPs and XPs for B\textmu's crossing the SSS close to its edge, as in this case both spacial and temporal alignment of the penetration points is difficult. Therefore, the upward-going contribution is reduced to 1-2\,\% if only IV\textmu's are regarded.

\btab
\captab{TraSig}{Percentage of mis-reconstructed (upward-going) muons for all ID\textmu's and IV\textmu's only.}
\begin{tabular}{|l|rr|}
\hline
		& ID\textmu's & IV\textmu's\\ 
\hline
OD tracking	& 2.0\%	& 1.2\% \\
ID tracking	& 5.2\%	& 0.8\% \\
GL tracking	& 3.2\%	& 2.1\% \\
\hline
\end{tabular}
\etab

\subsubsection{Cosmogenic Neutrons}

Cosmogenic neutrons are visible in Borexino via the $\gamma$-rays produced after their capture as explained in more details in \sec{neu}. 
The analysis of the distance $x$ of cosmogenic neutrons to their parent muon tracks is interesting for two reasons. First, the distance distribution of a large neutron sample can be used to determine the  mean free path of a cosmic neutron in the scintillator. This study is in preparation and will be presented as a separate paper. Second, for short distances, the $x$ distribution provides information on the precision of the track reconstruction. We focus here on this second outcome of the analysis.

\sfig{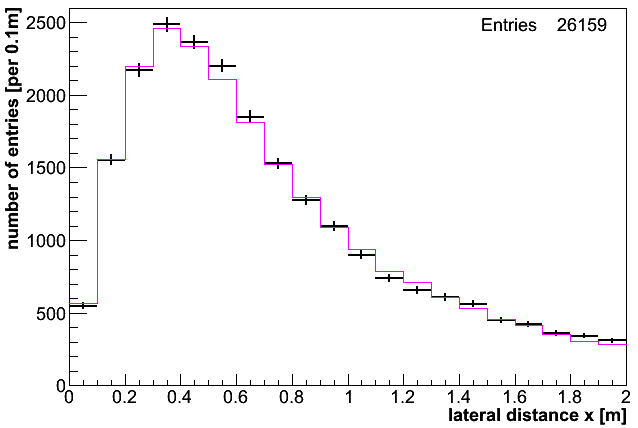}{.6}{Distribution of the distance $x$ of neutrons to their parent muons as derived from the global tracking. Crosses represent data points, the line corresponds to the MC result.}

\Fig{mune_distance} shows the $x$ distribution derived from global tracking: the neutron sample was selected requiring a minimum visible energy of 700 hits, corresponding to neutron capture events on hydrogen inside the FV (\fig{cnpmts}). This minimizes the effect of spatial uncertainty introduced by the reconstruction of the neutron capture vertex but also distorts the information on the mean free path of the neutrons by selecting only non-saturating (i.\,e.\,non-showering) muons, as explained in more detail in \sec{neu_pos}. 

This distribution can be reproduced by a Monte Carlo simulation that takes into account the muon lateral resolution
$\sigma_\mu$, the neutron mean free path $\lambda$, and the uncertainty in the reconstruction of the light barycenter of the 2.2\,MeV $\gamma$-rays ($\sigma_n$).
The muon lateral resolution function is composed of a Gaussian uncertainty $\sigma$ and a constant asymmetric offset $\mu$
that can be associated with radial shifts of the reconstructed tracks. For the neutron, $\sigma_n$ is dominated by the
reconstruction uncertainty and mean free path of the $\gamma$-ray ($\sigma_\gamma\approx23$\,cm was obtained by simulations), but the diffusion of the thermalized neutron before capture $\sigma_\mathrm{diff}\sim4$\,cm is also considered \cite{hoc06}. The best agreement to measured Borexino data is obtained by varying the parameters $\sigma$, $\mu$, $\lambda$ and the overall normalization in the MC distribution. The best fit values are listed in \tab{TabTraDist} and the corresponding MC fit for the global tracking is shown in \fig{mune_distance}.

The lateral resolution $\sigma_\mathrm{lat}$ is derived from the muon resolution parameters $\sigma$ and $\mu$ by calculating the root mean square of the corresponding distributions of lateral deviations relative to 0, i.e. the average lateral distance to the real track. The resulting values for $\sigma_\mathrm{lat}$ range from 30\,cm for global tracking to $\sim$40\,cm for OD and ID. This is comparable to the results derived from OPERA tracking. The fit value for the neutron mean free path $\lambda$ is close to earlier LVD results \cite{lvd99}. 

\btab
\captab{TabTraDist}{Best fit parameters from the numerical fit to the muon-neutron distance ($x$) distribution.}
\begin{tabular}{|l|cc|c|c|c|}
\hline
          		& $\sigma$\,[cm] & $\mu$\,[cm] & $\sigma_\mathrm{lat}$\,[cm] & $\lambda$\,[cm] & $\chi^2/\mathrm{ndf}$  \\
\hline
OD tracking	& 34.4$\pm$0.8 & 20.9$\pm$1.0 & 40.3$\pm$1.2 & 77.4$\pm$0.9 & 21.4/16 \\
ID tracking	& 34.2$\pm$0.8 & 21.0$\pm$0.9 & 40.2$\pm$1.2 & 73.9$\pm$0.9 & 25.9/16 \\
GL tracking	& 18.7$\pm$1.0 & 23.0$\pm$0.6 & 30.0$\pm$1.0 & 76.6$\pm$0.8 & 31.9/16 \\
\hline
\end{tabular}
\etab


\section{Muon-Induced Neutrons}
\label{sec:neu}

Cosmic muons crossing the SSS can produce neutrons through various spallation processes \cite{gal04} on carbon nuclei.
The scattering and thermalization of the neutron is invisible in Borexino as it cannot be disentangled from the much larger light emission of the muon.
However, the neutrons are subsequently captured in the detector with the emission of a 2.2\,MeV $\gamma$-ray 
if captured on hydrogen H or a 4.9\,MeV $\gamma$-ray if captured on carbon C. 
The typical capture time is of the order of 250\,\textmu s.
Detection of such $\gamma$-rays with the highest possible efficiency is of primary importance for Borexino 
both for antineutrino detection (\cite{bx10geo,bx11anu}) and for the rejection of cosmogenic isotopes produced 
in association with the neutrons themselves. 

\subsection{Neutron detection principles}
\label{sec:neu_det}

\sfig{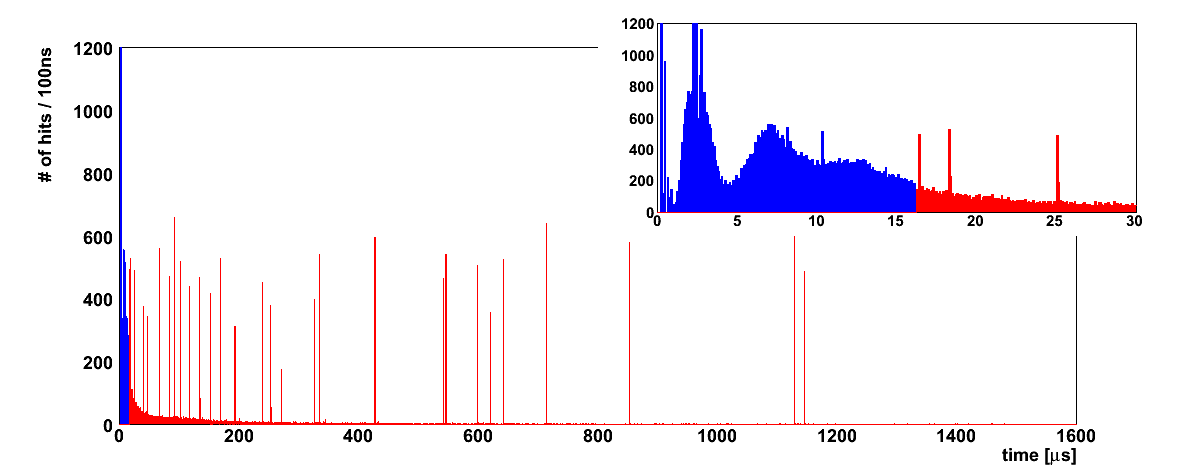}{1}{Hit time distribution for a sample IV\textmu\ event (blue) and the associated neutron detection gate (red). 53 neutron captures (spikes) were identified.
The smaller graph is a close-up view of the first 30\,$\mu s$.}

The difficulty for neutron detection arises from the muon after-pulse hits that are generated in the PMTs for several tens of \textmu s after the muon has already left the detector. 
This is particularly true for IV\textmu's, while it is a secondary problem for B\textmu's.  
The number of neutrons generated and captured in the active volume of the detector after a single muon event can extend up to $\sim$200, introducing further difficulty.
The challenging task of recording the highest possible number of $\gamma$-rays within a few tens of \textmu s is met as follows.
Defining \emph{muon} event an ordinary ID event flagged also by the OD (MTF flag), 
we programmed the trigger to generate after each \emph{muon} event a special \emph{neutron} event whose gate is 1.6\,ms, 
i.e. 100 times longer than an ordinary event. 
Although the \emph{muon} and \emph{neutron} events are meant to be consecutive a tiny gap could not be avoided and has been measured to be $(150\pm50)$\,ns.
Hits recorded in the \emph{neutron} event include the tail of the muon after-pulse hits distribution as well as the hits due to neutron captures if present. 
A sample pair of events where a high number of neutron captures is present can be seen in \fig{mu_plus_neutrons}.
A dedicated algorithm has been developed to search and isolate cluster of hits due to $\gamma$-rays 
that stand out above the "noise" background of after-pulse hits. 

Physical events not correlated with the muon may accidentally fall inside the long neutron gate and be identified as clusters by the algorythm.
They are generally due to $^{14}$C decays corresponding to about 50-100 hits and can be confused with partially contained $\gamma$-rays.
However, such clusters generally occur in single-cluster events and are mostly rejected by a proper cut introducing a negligible loss of signal.

To test the neutron detection efficiency, a dedicated hardware system has been developed and installed:
a 500\,MHz waveform digitizer (CAEN v1731) is fed with the analog sum of all ID PMTs properly attenuated.
A test performed based on a few months of data returns a neutron detection efficiency of $\sim$99\% for neutrons produced by ID\textmu's.

\subsection{Neutron capture spectrum}
\label{sec:neu_spectrum}

The energy spectrum for neutron capture $\gamma$-rays inside the fiducial volume FV (the innermost volume, a sphere of radius of 3.1m) is shown in \fig{cnpmts}.
The dashed line is the distribution of the number of hit PMTs, $n_\mathrm{h}$, normalized to 2000 live channels and shows two prominent features around 750 and 1250, due to $\gamma$-rays from capture on H and C respectively. 
The solid line is the same quantity corrected taking into account the binomial probability of multiple hits: 
\beq
N_\mathrm{corr} = -2000 \cdot \ln{\left(1 - \frac{n_\mathrm{h}}{f_\mathrm{ne} N_\mathrm{PMTs}}\right)}
\label{eq:n_corr}
\eeq
where $N_\mathrm{PMTs}$ is the number of working PMTs in the run and $f_\mathrm{ne}$ is the fraction of non saturated electronic boards in the specific event. 
This last correction is necessary because muons that cross several meters of scintillator can generate enough light to flood the memory of a fraction of the digital electronic boards, which then remain unavailable during the following neutron detection gate.
However, the number of such boards is recorded and the effect can be corrected as in \eq{n_corr}.

Moreover, neutrons whose production and capture point lie on different sides of the inner vessel (either way) can result in partially contained $\gamma$-rays.
For this reason the distribution of detected hits for neutron capture $\gamma$-rays extends down to below the detection threshold visible in \fig{cnpmts}. 
The structure around 200 hit PMTs is mostly due to neutrons occurring in events with many neutrons, induced by showering muons; 
the neutron visible energy is suppressed by the mechanisms described in \sec{neu_pos}.

\dfigpr{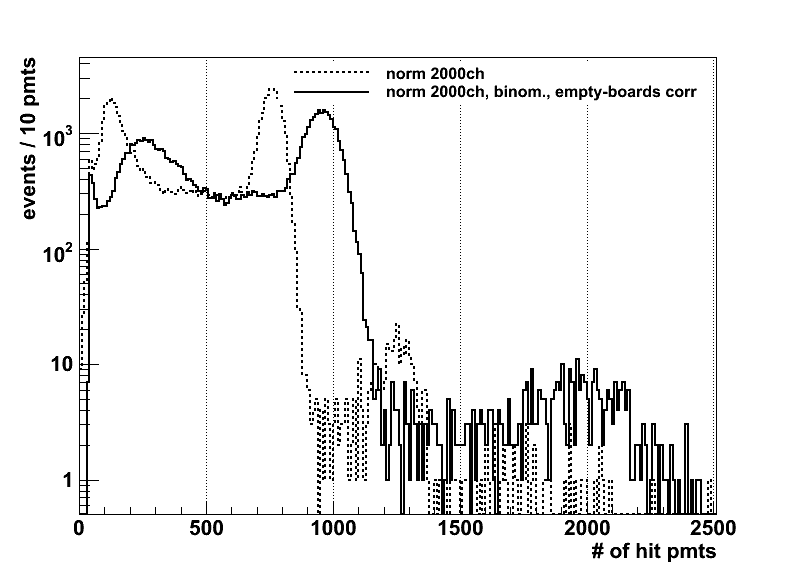}{Neutron capture spectrum (see \sec{neu_spectrum}).}
       {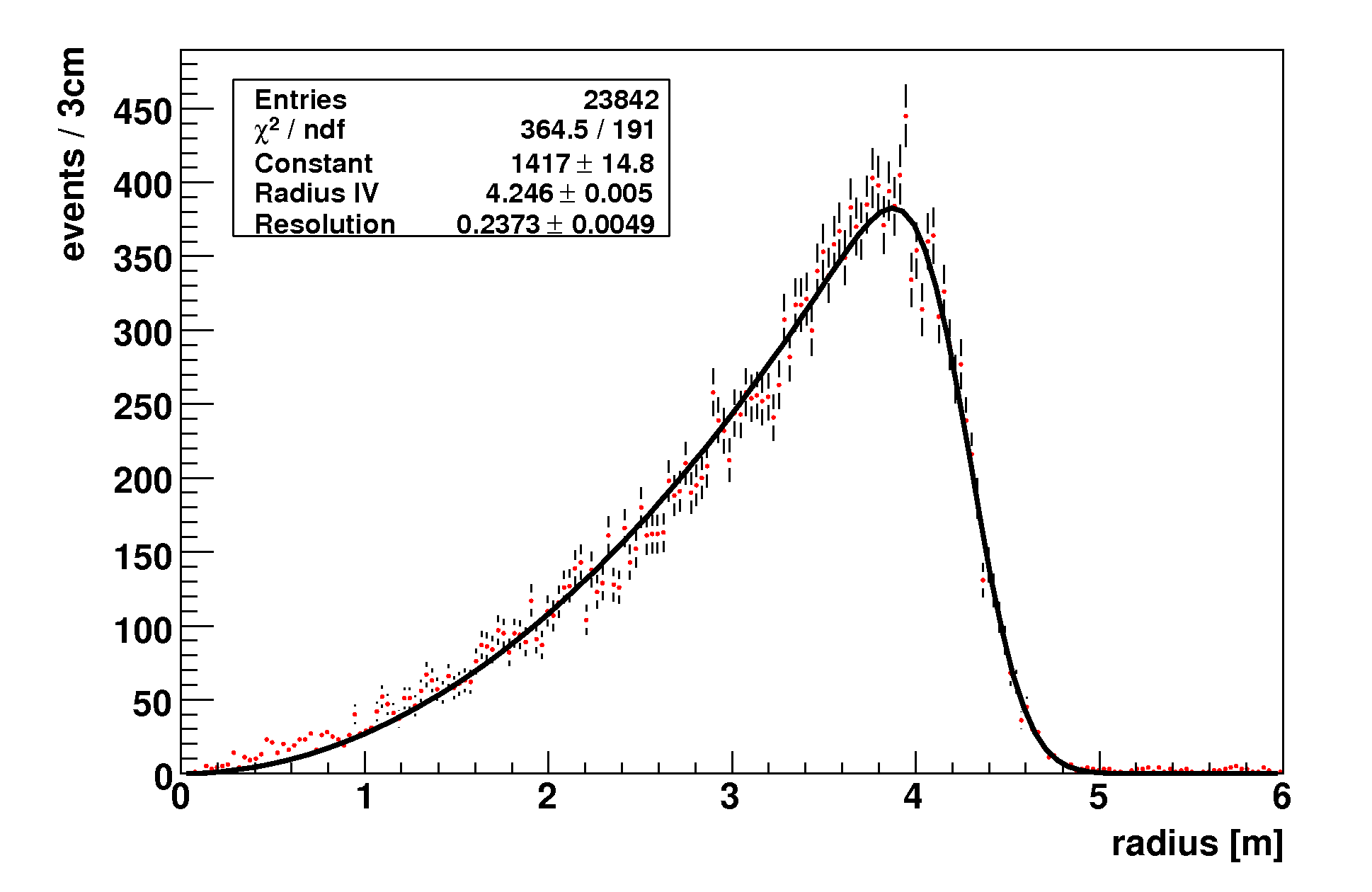}{Neutron radial distribution (see \sec{neu_pos}).}

\subsection{Neutron position reconstruction}
\label{sec:neu_pos}

The position reconstruction algorithm works well for neutron clusters belonging to muon events, that are not showering and generate normally a lower number of neutrons in the detector. This is the case for both B\textmu's and IV\textmu's. In general only a slight systematic shift of the reconstructed neutron positions to the center of the detector is observed.

On the contrary, muon events, that are showering and that generate a large number of neutrons, produce also a large amount of scintillation light.
This leads to two effects, which deteriorate the efficiency of the position reconstruction algorithm. 

First, large afterpulses are generated. In the time profile of the signal (\fig{mu_plus_neutrons}), the neutron capture peaks are superimposed on the tail of the afterpulse hits, affecting the position reconstruction algorithm as there is no way to discriminate $\gamma$-ray hits from after-pulse hits.

The second effect is that a fraction of the ID electronics boards can be saturated  and become unavailable for neutron detection (as mentioned before). Since PMTs were cabled in bundles, the saturated boards correspond to PMTs preferentially located on the half detector closer to the muon track. This decrease of detector uniformity aggravates the effect of afterpulse hits on the position reconstruction. To overcome this problem and to improve the position reconstruction of neutrons, we have developed a new custom firmware for the digital electronics that accepts only 25 hits per channel in the muon event, thus preventing the saturation of the memory in the neutron event. Data acquired from middle 2010 onwards will benefit from this improvement.

\Fig{radial} shows the radial distribution of neutron capture events. 
Here we have selected only neutron captures past the first 300 \textmu s after the muon, to limit the influence of after-pulse, and above an energy threshold of 200 hits (or lower depending on the level of empty boards), to exclude accidental coincidences due to {$^{14}$C}.
The distribution was fitted by the convolution of a volumetric function with a Rayleigh function that describes the detector resolution. 
The best fit returns $\chi^2/\mathrm{ndf}=364/191$. 
The goodness of the fit is limited by the fact that a higher energy threshold progressively suppresses neutron captures occurring close to the IV surface whose $\gamma$-rays are only partially contained and can introduce distortions in the radial profile at large radii.

\subsection{Neutron capture time}
\label{sec:neu_time}

\dfigsr{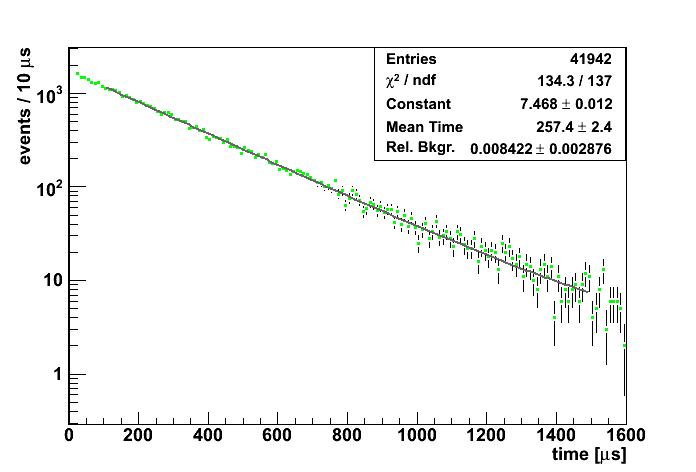}{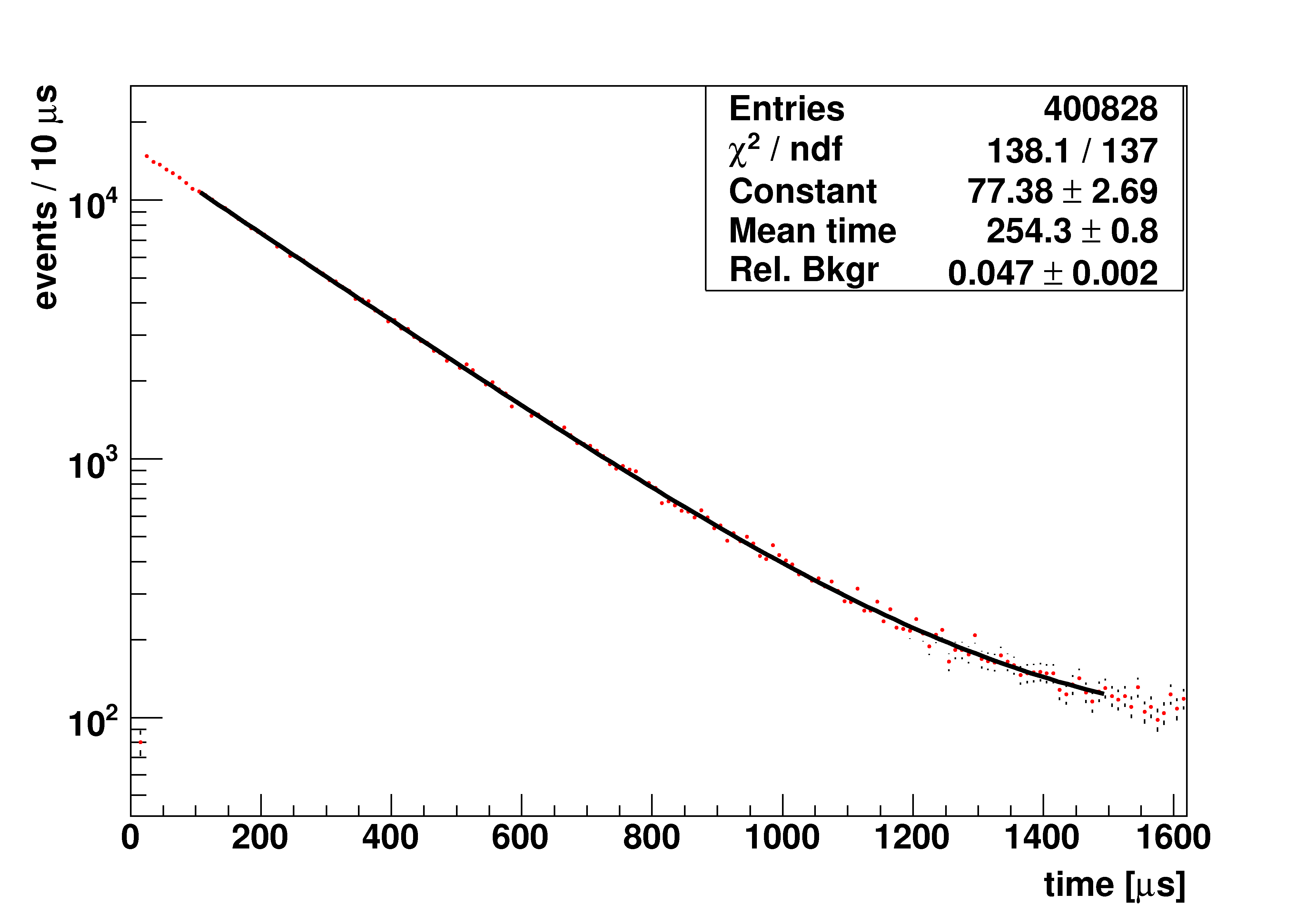}{capture_time}{Neutron capture time from cosmogenic neutrons (left panel) and from the Am-Be source campaign (right panel).}

Neutrons produced in muon interactions are slowed down to sub-eV energies in a few scattering interactions and this typically occurs within a few tens of ns. 
Thereafter, the thermal neutron can wander in the detector for a relatively long time before capture on H or C occurs. 
The capture time profile shows an exponential behavior (\fig{capture_time}, left panel) and depends mainly on the H density of the medium.
The fit is performed based on an exponential plus a constant term in the [100,1500]\,\textmu s range where the neutron detection
efficiency is believed to be uniform. The fit is excellent ($\chi^2/\mathrm{ndf}=134/137$) and returns $\tau = (257.4 \pm 2.4)$\,\textmu s and a background component in the sample of about 0.8\% (mostly $^{14}$C). 
However the results are not very stable against radial selection of the active volume and against an energy cut, due to the combined effects of the muon after pulse. 
A conservative estimation of the systematical error is about $\pm 8$\,\textmu s.

A far more accurate measurement of the neutron capture time has been performed by studying an Am-Be neutron calibration source which was inserted into the detector and deployed in more than 20 positions within the Inner Vessel \cite{har10phd}. The capture time was
determined by the selection of coincidences between prompt and delayed signals from the source. The maximal distance of correlated events was set to 2\,m in order to contain most of the emitted neutrons. The fit of the time profile of the source in the center of the
detector (\fig{capture_time}, right panel) returns $\tau = (254.3 \pm 0.8)$\,\textmu s ($\chi^2/\mathrm{ndf}=138/137$). 
The mean neutron capture time obtained from several source positions with up to 4.0 m distance from center of the detector is
$\tau = (254.5 \pm 1.8)$\,\textmu s. 

\section{Conclusion}
\label{sec:conclusion}

A  key element for the underground Borexino experiment is the suppression of the background induced by the residual cosmic muons.
A high-efficiency veto and geometrical reconstruction of these muons is mandatory for an identification of cosmogenic backgrounds based on time and space coincidences with the parent muon.

The Borexino detector is formed by an inner detector which contains liquid scintillator, the central target volume surrounded by a buffer. 
It is also equipped with an outer detector, a water tank, which serves both as an additional passive shielding against external radiation and as an active veto, identifying crossing muons via their \v{C}erenkov light emission.
The inner detector has been largely described in \cite{bx08det}. 
The components and characteristics of the outer detector have been detailed in the present paper. 

The identification of muons is based on the complementarity between outer and inner detectors.
Different methods of muon identification have been studied and their veto efficiencies calculated. 
The overall Borexino muon identification efficiency reaches 99.992\,\%. 

The reconstruction algorithms for muon tracks have been described and their performances studied using different methods. 
The resolutions achieved are at the level of $\sim$$3.5^\circ-5^\circ$ angular and $\sim$35-50\,cm lateral.   

The cosmogenic neutrons generated by muons have been studied and special techniques have been developed to increase the efficiency for vetoing both neutrons and subsequent decays of cosmogenic radioisotopes.

The study of cosmogenic isotopes induced by muons will be performed in a forthcoming paper.

\hspace{10mm}

{\bf Acknowledgments}

We thank the funding agencies: INFN (Italy), NSF (USA), BMBF, DFG and MPG (Germany), Rosnauka (Russia), MNiSW (Poland), and we acknowledge the generous support of the Laboratori Nazionali del Gran Sasso. This work was also partially supported by PRIN JR4STW 2007 by MIUR, Italy.
We thank Dario Autiero (IN2P3-Lyon)  for kindly providing us with  information about the CERN CNGS gateway Data Base\,; we also thank Paola Sala (INFN - Milano) and Alberto Guglielmi (INFN - Padova) for information about their CNGS Monte Carlo results.
We thank Mark Chen, who contributed to the early phases of this study. We thank for the design and construction of the Muon Veto electronics C. Gomez, J. Fitch, M. Pleasko, and B. Wadsworth of MIT, C. Sule of Princeton University, N. Frank and W. Liebl of TUM. We wish to express our gratitude to the OPERA Collaboration for providing us with the tracking information of the sample used in \sec{perform_cngs}.

\appendix

\section{The CNGS neutrino beam in Borexino}
\label{app:cngs}

The CERN to Gran Sasso (CNGS) neutrino beam \cite{web-proj-cngs,Acquistapace} produces muons 
in the rock upstream of Borexino via charged current interactions of the $\nu_{\text{\textmu}}$s. 
Further charged and neutral current $\nu_{\text{\textmu}}$ interactions occur inside the detector. 
These events are a background to neutrino analysis and must be removed from the data set. 
At the same time they constitute a sample of muons of known direction 
and we use them to determine Borexino muon tagging efficiency as explained in \sec{VetCNGS}. 
In addition, since Borexino is just in front of the OPERA detector in Hall C, 
we exploited the on-time events common to the two experiments to test the Borexino tracking algorithms 
for horizontal tracks (\sec{perform_cngs}).

The mean neutrino energy is about 18\,GeV and the mean energy of the induced muons is about 10\,GeV \cite{Ferrari-rep}.
The expected number of muons per charged current neutrino interaction is one, 
with only 0.3 $\%$ of the interactions producing more than one muon. 
At the LNGS, the neutrino beam arrives with an incline of 3.2$^\circ$ below the horizon, 
corresponding to a a zenith angle of $93.2^\circ$; the azimuth angle with respect to Hall B axis 
(parallel to Hall C, where Borexino is located) 
has been measured to be $0.18^\circ$ \cite{Dario-rep}; 
its width is more than one kilometer.

For synchronization purposes, both CERN and LNGS have timing system based on GPS, 
with comparable resolution ($<$ 100 ns) and accurately intercalibrated \cite{Serrano:2006zz}. 
The time delays between the LNGS master GPS receiver located above ground 
and the slave units located at each underground experiment have been accurately measured. 
In Borexino the correlation with the beam extraction windows can be obtained by comparing 
the GPS time of each trigger with the ones obtained from the CERN database. 
The time-of-flight distribution shown in \Fig{diff_all_and_uff} (left) features 
a clear peak at 2.4\,ms, the expected value for the distance from CERN to LNGS. 
\Fig{diff_all_and_uff} (right) shows the same quantity on a finer scale allowing 
to see the $\sim$10\,\textmu s spread that corresponds to the extraction time duration.
The events in the time window [2.40, 2.416] ms are flagged as ``CNGS on-time events''. 

The fraction of accidental coincidences falling in this window 
-- mostly due to $^{14}$C decays in the scintillator -- can be easily calculated
from the average trigger rate, the number of on-time events and the number of extractions during Borexino uptime derived from known beam parameters.
We obtain $f\sim$1.9\,\% (3.6\,\%) if all (only ID) events are considered.
This reduces to below $\sim$0.06\,\% if only events above 80 hits are considered as in \sec{VetCNGS}.

\sfig{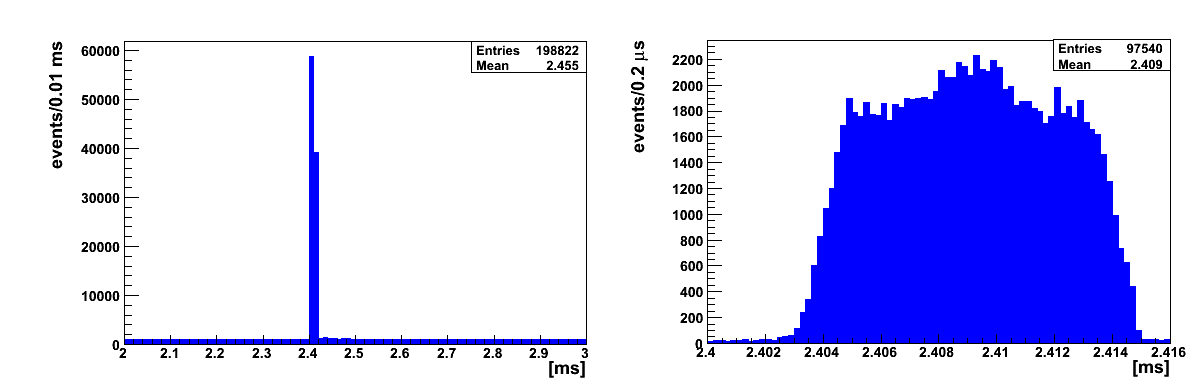}{1.}{CERN-Borexino time-of-flight from clock time differences.}

\Tab{pot} summarizes for each year of CNGS operation the beam intensity and the number of on-time events in Borexino.   
Since some of our calibration campaigns took place during beam time, 
the integrated intensity in the Borexino data sample is generally lower than the intensity delivered from CERN.

\btab
\captab{pot}{CNGS beam operation periods; Integrated beam intensity delivered by CERN; integrated beam intensity in the Borexino data sample; 
on-time events detected by Borexino; on-time ID events. Intensities are given in units of $10^{19}$ p.o.t. (protons on target). }
\begin{tabular}{|crcl|cc|cc|}
\hline
year   & \multicolumn{3}{c|}{beam period}  & $I$ delivered    & $I$ in BX data   & on-time & ID on-time \cr
       &        & &                        & $10^{19}$ p.o.t. & $10^{19}$ p.o.t. & events  & events     \cr 
\hline
2007   & Sep 22 &-& Oct 20                 & 0.08	      &  0.07 	         &     873 &    438     \cr
2008   & Jun 18 &-& Nov 03                 & 1.78             &  1.33            &   17687 &   9362     \cr
2009   & May 27 &-& Nov 23                 & 3.52             &  2.30            &   33206 &  17765     \cr
2010   & Apr 20 &-& Nov 22                 & 4.04             &  3.16            &   45774 &  24749     \cr
\hline
\multicolumn{4}{|c|}{total}                & 9.42             &  6.86            &   97540 &  52314     \cr
\hline
\end{tabular}
\etab


\end{document}

%% file: defines.tex
\newcommand{\bit}{\begin{itemize}}
\newcommand{\eit}{\end{itemize}}
\newcommand{\ben}{\begin{enumerate}}
\newcommand{\een}{\end{enumerate}}
\newcommand{\bde}{\begin{description}}
\newcommand{\ede}{\end{description}}
\newcommand{\bqu}{\begin{quote}}
\newcommand{\equ}{\end{quote}}
\newcommand{\bmp}[1]{\begin{minipage}[t]{#1\textwidth}\centering}
\newcommand{\emp}{\end{minipage}}

\newcommand{\beq}{\begin{equation}}
\newcommand{\eeq}{\end{equation}}
\newcommand{\bea}{\begin{eqnarray}}
\newcommand{\eea}{\end{eqnarray}}
\newcommand{\bdm}{\begin{displaymath}}
\newcommand{\edm}{\end{displaymath}}

\makeatletter
  \newcommand{\capfig}[2]{\def\@captype{figure}\caption[]{\label{fig:#1} #2}}
  \newcommand{\captab}[2]{\def\@captype{table}\caption[]{\label{tab:#1} #2}}
\makeatother

\newcommand{\bfig}{\begin{figure}[htb] \centering}
\newcommand{\bottomfig}{\begin{figure}[b] \centering}
\newcommand{\efig}{\end{figure}}
\newcommand{\btab}{\begin{table}[htb] \centering}
\newcommand{\etab}{\end{table}}
\newcommand{\bff}[2]{\begin{floatingfigure}[#1]{#2\textwidth}}
\newcommand{\eff}{\end{floatingfigure}}

\newcommand{\getfigw}[2]{\includegraphics[width=#2\textwidth]{#1.png}}

\newcommand{\getfigr}[3]{\includegraphics[width=#2\textwidth, height=#3\textwidth]{#1.png}}

\newcommand{\sfig}[3]{\bfig \getfigw{#1}{#2}  \capfig{#1}{#3} \efig}

\newcommand{\dfigsr}[4]{\bfig \getfigr{#1}{0.49}{.3369} \hfill \getfigr{#2}{0.49}{.3369}  \capfig{#3}{#4} \efig}

\newcommand{\dfigp}[6]{\bfig \bmp{#2} \getfigw{#1}{1} \capfig{#1}{#3} \emp \hfill \bmp{#5} \getfigw{#4}{1} \capfig{#4}{#6} \emp \efig}

\newcommand{\dfigpr}[4]{\bfig \bmp{0.49} \getfigr{#1}{1}{0.6875} \capfig{#1}{#2} \emp \hfill \bmp{0.49} \getfigr{#3}{1}{0.6875}\capfig{#3}{#4} \emp \efig}

\newcommand{\fig}[1]{fig.~\ref{fig:#1}}
\newcommand{\Fig}[1]{Fig.~\ref{fig:#1}}
\newcommand{\tab}[1]{tab.~\ref{tab:#1}}
\newcommand{\Tab}[1]{Tab.~\ref{tab:#1}}
\newcommand{\eq}[1]{eq.~\ref{eq:#1}}

\renewcommand{\sec}[1]{sec.~\ref{sec:#1}}
\newcommand{\Sec}[1]{Sec.~\ref{sec:#1}}

\newcommand{\app}[1]{app.~\ref{app:#1}}

\newcommand{\com}[1]{\textsc{\large{#1}}}
\newcommand{\che}{\v{C}erenkov}
\newcommand{\imply}{\ensuremath{\qquad\Rightarrow\qquad}}